\documentclass[thmsa,a4paper,12pt]{article}
\usepackage{enotez}
\usepackage{graphicx,psfrag,epsf}
\usepackage{enumerate}
\usepackage{amsmath,amssymb,amsthm}
\usepackage{natbib}
\usepackage{color}
\usepackage[normalem]{ulem}
\usepackage{caption}
\usepackage{tabularx}
\usepackage{float}
\usepackage{morefloats}
\usepackage{lscape}
\usepackage{rotating}

\RequirePackage[OT1]{fontenc}
\RequirePackage[colorlinks,citecolor=blue,urlcolor=blue]{hyperref}

\def\tecb{\textcolor{blue}}

\newtheorem{theorem}{Theorem}
\newtheorem{proposition}{Proposition}
\newtheorem{lemma}{Lemma}

\newtheorem{Acondition}{}

\newtheorem{Bcondition}{}

\def\plim{\mathop {\rm plim}}

\def\cov{\hbox{cov}}
\def\and{\mbox{ and }}

\def\ba{\begin{array}}
\def\bc{\begin{center}}
\def\bd{\begin{description}}
\def\be{\begin{enumerate}}
\def\ec{\end{center}}
\def\ed{\end{description}}
\def\ee{\end{enumerate}}
\def\ea{\end{array}}
\def\ben{\begin{equation}}
\def\benn{\begin{equation*}}
\def\een{\end{equation}}
\def\eenn{\end{equation*}}
\def\benr{\begin{eqnarray}}
\def\eenr{\end{eqnarray}}
\def\benrr{\begin{eqnarray*}}
\def\eenrr{\end{eqnarray*}}
\def\al{\alpha}

\def\b{\beta}

\def\del{\delta}

\def\edt{\end{document}}
\def\ep{\epsilon}

\def\g{\gamma}

\def\h{\hat}

\def\iny{\infty}
\def\ka{\kappa}

\def\la{\lambda}

\def\mb{\mbox}
\def\noi{\noindent}

\def\nn{\nonumber}
\def\np{\newpage}

\def\om{\omega}

\def\r{\ref}
\def\R{{\mathbb R}}

\def\ra{\rightarrow}

\def\s{\sum_{i=1}^n}

\def\Si{\Sigma}

\def\vep{\varepsilon}

\def\vs{\vskip}

\def\wh{\widehat}

\def\Z{{\mathbb Z}}

\def\bfdelta{{\boldsymbol{\delta}} }
\def\bftheta{{\boldsymbol{\theta}} }

\def\bfphi{{\boldsymbol{\phi}} }

\def\cA{{\mathcal{A}}}
\def\cB{{\mathcal{B}}}
\def\cC{{\mathcal{C}}}
\def\cD{{\mathcal{D}}}

\def\cF{{\mathcal{F}}}

\def\cH{{\mathcal{H}}}

\def\cM{{\mathcal{M}}}
\def\cN{{\mathcal{N}}}

\def\cU{{\mathcal{U}}}

\def\var{\hbox{Var}}

\def\mE{{\mathrm{E}}}

\def\vec#1{\oalign{#1\crcr\hidewidth \vbox to .3ex{\kern 1pt
           \hbox spread .1em
           {\hss\char'176\hss}\vss}\hidewidth}}
\def\bfA{{\boldsymbol{A}}}

\def\bfI{{\boldsymbol{I}}}

\def\bfU{{\boldsymbol{U}}}

\def\bfa{{\boldsymbol{a}}}
\def\bfb{{\boldsymbol{b}}}

\def\bfs{{\boldsymbol{s}}}

\input{tcilatex}

\begin{document}

\title{ {\Large Specification tests for
GARCH processes}}
\author{ Giuseppe Cavaliere\thanks{
Department of Economics, University of Exeter, UK and Department of Economics, University of Bologna, Italy \
(Email: giuseppe.cavaliere@unibo.it) }, \ Indeewara Perera\thanks{
Department of Economics, The University of Sheffield, 
UK
\
(Email: i.perera@sheffield.ac.uk) } \ and Anders Rahbek\thanks{
Department of Economics, University of Copenhagen, Denmark \
(Email: anders.rahbek@econ.ku.dk)} }
\maketitle

\baselineskip 17.5pt

{
\renewcommand{\thefootnote}{}
\setcounter{footnote}{1}
\footnotetext{This research was supported by the Danish Council for Independent Research (DSF Grant 015-00028B) and by the Italian
Ministry of University and Research (PRIN 2017 Grant 2017TA7TYC).}
\setcounter{footnote}{0}
}

\renewcommand{\thefootnote}{\arabic{footnote}}

\begin{abstract}
This paper develops tests for the correct specification of the conditional
variance function in GARCH models when the true parameter may lie on the
boundary of the parameter space. The test statistics considered are of
Kolmogorov-Smirnov and Cram\'{e}r-von Mises type, and are based on a certain
empirical process marked by centered squared residuals. The limiting
distributions of the test statistics are not free from (unknown) nuisance
parameters, and hence critical values cannot be tabulated. A novel bootstrap
procedure is proposed to implement the tests; it is shown to be
asymptotically valid under general conditions, irrespective of the presence
of nuisance parameters on the boundary. The proposed bootstrap approach is
based on shrinking of the parameter estimates used to generate the bootstrap
sample toward the boundary of the parameter space at a proper rate. It is
simple to implement and fast in applications, as the associated test
statistics have simple closed form expressions. A simulation study
demonstrates that the new tests: (i)\ have excellent finite sample behaviour
in terms of empirical rejection probabilities under the null as well as
under the alternative; (ii)\ provide a useful complement to existing
procedures based on Ljung-Box type approaches. Two data examples are
considered to illustrate the tests.
\end{abstract}

\vs .25in




\np

\section{Introduction}

Generalized autoregressive conditionally heteroskedastic (GARCH) models introduced by \cite{Bollerslev:86}
are widely used for modelling various financial time series processes. 
The data generation mechanism of a GARCH model requires the
conditional variance to be always strictly positive, which is generally obtained by imposing
a strictly positive intercept and non-negative GARCH coefficients in the conditional variance
equation. Consequently, in GARCH models, the admissible
parameter space typically needs to be inequality restricted.
This represents an important difference between
GARCH and other popular time series models, such as AR and ARMA models.
Although omnibus specification testing in GARCH type models against unspecified alternatives has attracted considerable attention in the recent literature,
a crucial weakness in the current theory remains the exclusion of the presence of nuisance parameters on the boundary.
This paper contributes towards
addressing this issue by developing new statistical methodology for specification testing in GARCH models.
%

There are a number of different GARCH models available in the literature and many of them are nonnested models (see \citealp{Francq:2010}).
Therefore, in many cases, a sensible way to proceed when testing a specification of a GARCH model is to leave the alternative model unspecified, or to test the lack-of-fit.
This type of tests, also known as omnibus tests, have their roots in the seminal work of \cite{Kolmogorov:33} on testing for a specific probability distribution function, and
\cite{Grenander:Rosenblatt:57} on testing the hypothesis of white noise dependence.
Several omnibus specification tests in GARCH type models 
have been proposed in the literature.
These include tests based on weighted empirical processes of standardized residuals (\citealp{Koul:06,Escanciano:2010}),
spectral distributions based tests in the frequency domain (\citealp{Hidalgo:Zaffaroni:07,Escanciano:2008}),
residual based tests for nonnegative valued processes (\citealp{Fernandes:Gramg:05,Koul:Indee:12,Perera:Silvapulle:14}),
and Khmaladze type (\citealp{khm:81}) martingale transformations based tests (\citealt{Bai:03,Indee:Hira:15}), amongst~others. 

A key regularity condition imposed by the aforementioned
specification tests 
is to restrict the true parameter to the interior of the null parameter space.
%
Since the parameter space of a GARCH-type model is inequality restricted,
this condition is not typically satisfied if some ARCH or GARCH coefficients are zero,
 because then the true parameter may lie on the boundary of the parameter space.
 Therefore, for the theory developed in the above cited papers,
the true parameter being an interior point is essential;
for example, the limiting process obtained
in Theorem~2.1 in~\cite{Hidalgo:Zaffaroni:07} would not be Gaussian if, for instance, a GARCH($p, q$) model is
estimated when the underlying true process is a GARCH($p-1, q$), or a GARCH($p, q-1$) process.
Similarly, 
the asymptotic properties of the other aforementioned papers would also not hold when some nuisance parameters lie on the boundary.

In this paper we contribute towards the literature of specification testing in GARCH models by developing a new class of tests for the correct specification of the conditional variance function
while allowing the null model to have an unknown number of nuisance parameters on the boundary of the parameter space.
Our test statistics are 
functionals of an
empirical process marked by centered squared residuals and are easy to compute. The limiting distributions of the test statistics are not free from (unknown) nuisance parameters, and hence critical values cannot be tabulated for general use. We propose a bootstrap method to implement the tests and show that it is asymptotically valid under general conditions, irrespective of the presence of nuisance parameters on the boundary. The proposed bootstrap approach is simple to implement, and is based on a method of shrinkage of the parameter estimates used to generate the bootstrap sample toward the boundary of the parameter space at an appropriate rate.
This approach is similar to the related bootstrap scheme 
advocated in \cite{Cavaliere:19}, in a different context,
for bootstrapping likelihood ratio statistics,
and it also has its roots in the modified bootstrap approach considered in \cite{Chatterjee:Lahiri:11} for bootstrapping Lasso-type estimators.
Our bootstrap tests are shown to be consistent against fixed alternatives.
We also separately consider the case the nuisance parameters lie in the interior of the parameter space for Kolmogorov-Smirnov and Cram\'{e}r-von Mises type tests based on the aforementioned marked empirical process,
and show that the bootstrap implementations of these tests under standard residual based bootstrap are asymptotically valid and consistent.
Our tests can be implemented easily because the test statistics have simple closed form expressions.  A simulation study shows that the proposed tests have desirable finite sample properties.
We illustrate the testing procedure by considering two real data examples.

The rest of this paper is structured as follows.
Section~\ref{sec:mot:eg} formulates the problem, defines the estimators and test statistics.
Section~\ref{sec:3} provides the results relating to the asymptotic validity and consistency of the bootstrap tests
when the parameters are in the interior of the parameter space.
Section~\ref{sec:4} considers inference when some components of the true parameter lie on the boundary of the parameter space.
Section~\ref{Sim-boot-bound} describes a simulation study.
Two empirical illustrations are discussed in Section~\ref{sec:example}.
Section~\ref{sec:con} concludes the paper. The proofs and some assumptions are relegated to Appendix~\ref{appendixA}.


\section{Formulation of the Problem}\label{sec:mot:eg}

Let $(Y_1, Y_2, \ldots, Y_n)$ be a realization of an observable stationary process $\{Y_i\}$ satisfying
\ben\label{con:var:mod}
Y_i = {\it h}_i^{1/2}\vep_i, \quad i \in \Z :=\{0,\pm1,\pm 2,\cdots\},
\een
where the errors $\varepsilon_i,\, i\in\Z,$ are
independent and identically distributed (i.i.d.)
random variables (r.v.'s) having zero mean and unit variance
with common cumulative distribution function (c.d.f.) $F_0$,
and ${\it h}_i = \mE[Y_i^2 \mid \cH_{i-1}]$,
where $\cH_{i-1}$ denotes the information available up to time $i-1$
for forecasting $Y_i$, $i \in \Z$.

As is well-known, a GARCH($p_1,p_2$) model for ${\it h}_i$ takes
the form
\ben\label{vsi:mod}
{\it h}_i = {\it h}_i(\bfphi) = \om + \sum_{j=1}^{p_1} \al_{j} Y_{i-j}^2
+ \sum_{k=1}^{p_2} \b_{k} {\it h}_{i-k}(\bfphi), \quad i \in \Z;
\een
the vector of parameters $\bfphi = (\phi_1, \ldots, \phi_{p_1+p_2+1})^\prime = (\om, \al_1, \ldots, \al_{p_1}, \b_1, \ldots, \b_{p_2})^\prime$,
usually belongs to a compact parameter space
\ben\label{Phi}
\Phi \subset (0, \iny) \times [0, \iny)^{p_1+p_2}
\een
with
$\om > 0$, $\al_{k} \ge 0 \, (k=1,\ldots, p_1), \b_{k} \ge 0 \, (k=1, \ldots, p_2)$, and
in order to avoid well-known identification issues (see also Assumption~\ref{qml:4} below) one typically imposes
$\sum_{k=1}^{p_1} \al_{k} \ne 0$.

Suppose we wish to test the adequacy 
of the above GARCH($p_1,p_2$) model for ${\it h}_i$, i.e., to test the null hypothesis
\benr\label{null1}
\mathsf{H}_0:&&  {\it h}_i = {\it h}_i(\bfphi_0) = \om_0 + \sum_{j=1}^{p_1} \al_{0j} Y_{i-j}^2
+ \sum_{k=1}^{p_2} \b_{0k} {\it h}_{i-k}(\bfphi_0), \ \mbox{a.s.} \ \text{for all $i$, and}\\ %
\notag && \text{for some $\bfphi_0 = (\om_0, \al_{01}, \ldots, \al_{0p_1}, \b_{01}, \ldots, \b_{0p_2})^\prime \in \Phi$,}
\eenr
against the alternative \emph{$\mathsf{H}_1 : \mathsf{H}_0$ is not true.}

Since some ARCH or GARCH coefficients may be zero, %
the null model~\eqref{null1} allows some components of $\bfphi_0$ to be on the boundary of the parameter~space $\Phi$.

Let $\h \bfphi$ denote
the \textit{Gaussian quasi maximum likelihood estimator} [QMLE]
defined by
\ben\label{lth}
\h \bfphi = \arg\min_{\bfphi \in \Phi} \s \ell_i(\bfphi),
 \quad \ell_i(\bfphi) = \log {\it h}_i(\bfphi) + [Y_i^{2}/{{\it h}_i(\bfphi)}],
\een
with ${\it h}_i(\bfphi)$ being defined recursively by~\eqref{vsi:mod} for $i=1, 2, \ldots, n$.
To simplify the exposition, the vector of initial values,
$\varsigma_0=(Y_0, \ldots, Y_{1-p_1}, {\it h}_0, \ldots, {\it h}_{1-p_2})^\prime \in \R^{p_1} \times [0,\iny)^{p_2}$,
%
is assumed to be fixed for the statistical analysis. The asymptotic results do not change if $\varsigma _{0}$ is
replaced by an arbitrarily chosen vector (e.g., by setting $Y_{t}=0$ and ${\it h}_{t}=0$, all $t\leq 0$); see, for example,
the discussions in \cite{Straumann:Mikosch:06}, \cite{Perera:Silvapulle:20} and \cite{Jensen:Rahbek:04}.

Let 
$
\widehat{\varepsilon}_i  :=
Y_i/\{{\it h}_i(\hat{\bfphi})\}^{1/2},  \ i =1, \ldots, n,
$
denote the estimated residuals.
 With $\boldsymbol{\phi } \in \Phi,$ we propose an omnibus test statistic based on the
marked empirical~process:
\ben\label{eq1:0}
\mathcal{U}_n(y,\bfphi)
:= n^{-1/2} \sum_{i=1}^n \left\{\frac{Y_i^2}{{\it h}_i(\bfphi)} - 1\right\} \mathbb{I}(Y_{i-1} \leq y),
\quad y \in \R, \, \bfphi \in \Phi,
\een
where $\mathbb{I}$ denotes the indicator function.
We allow the domain of $\cU_n(\cdot, \bfphi)$ to extend over the whole real line by letting
$\cU_n(-\iny, \bfphi):=0$ and $\cU_n(\iny, \bfphi):=  n^{-1/2} \sum_{i=1}^n \{Y_i^2/{\it h}_i(\bfphi) - 1\}$. 
Hence, $\cU_n(\cdot, \bfphi)$  in~\eqref{eq1:0}
can be viewed as a process in the space of \textit{cadlag} functions on $[-\iny,\iny]$, equipped with the uniform~metric,
which we denote by $\cD(\R)$.
%
This process
is an extension of the so-called
cumulative sum process for the one sample setting to the current set up.
Under~$\mathsf{H}_0$, $\mE\cU_n(y,\bfphi_0) = 0$, for all $y$, but not under $\mathsf{H}_1$.
Hence, if $\mathsf{H}_0$ is true, then we would expect $\cU_n(y, \h \bfphi\,)$ to be close to zero for all $y$, but not otherwise. Therefore,
a suitable functional of $\cU_n(\cdot, \h \bfphi\,)$
can potentially be used as a test statistic for testing $\mathsf{H}_0$ against~$\mathsf{H}_1$.

The use of cumulative
sum processes for specification testing similar to $\cU_n(\cdot, \h \bfphi\,)$ 
goes back to \cite{Neumann:41}, who proposed a test of
constant regression based  on an analog of this process.
A motivation for basing inference in nonnegative valued processes on an analog of the process
$\mathcal{U}_n(\cdot,\bfphi), \, \bfphi \in \Phi,$ also appears in \cite{Koul:Indee:12}.
Similar tests have also been considered
by \cite{Stute:97} and \cite{Koul:Stute:99} for certain regression and additive time series models.
More recently, analogs of $\cU_n(\cdot, \h \bfphi\,)$
have been used by several authors to propose asymptotically distribution free specification tests
 in related time series models; see, for example, \cite{Indee:Hira:15} and \cite{Bala:Koul:19}.
 In the econometric analyses presented in these
 papers certain tests based on analogs of $\cU_n(\cdot, \h \bfphi\,)$ have demonstrated desirable finite sample and asymptotic properties.
Therefore, we find it of interest to develop specification tests based on similar
statistics involving the process $\cU_n(\cdot, \h \bfphi\,)$ for testing $\mathsf{H}_0$ against $\mathsf{H}_1$ in the current setup.
In particular, we consider~the Kolomogorov-Smirnov (KS) and Cram\'{e}r-von Mises (CvM) type statistics
which can be defined in terms of $\cU_n(\cdot, \h \bfphi)$ as:
\begin{equation}
\label{eq-T1}
T_1 := \mbox{KS}  =\sup_{y}\big|\cU_n(y, \h \bfphi)\big|, \quad
T_2 := \mbox{CvM}  = \int \cU_n^2(y, \h \bfphi\,) dG_n(y),
\end{equation}
where $G_n(y) := n^{-1}\s \mathbb{I}(Y_{i-1}\le y)$.
Other suitable functionals of $\cU_n(\cdot, \h \bfphi\,)$ may also be considered as possible test statistics (see \citealp{Stephens:86}).

\section{Inference when the parameters are in the interior of the parameter space}\label{sec:3}

Before moving to the general case which includes possible parameters on the boundary of the parameter space, we here consider the case the true parameter $\bfphi_0$ is in the interior of $\Phi$.

The asymptotic distribution of ${\mathcal{U}}_{n}(\cdot ,{\boldsymbol{\phi }}_{0})$
under the null hypothesis $\mathsf{H}_{0}$ can be derived by standard arguments,
under the assumptions on the GARCH process discussed in the next subsection.
Specifically, from a martingale central limit theorem [for example \cite%
{Hall:Heyde:80}, Corollary 3.1] and the Cram\'{e}r-Wold device it follows
that all finite dimensional distributions of ${\mathcal{U}}_{n}(\cdot ,{%
\boldsymbol{\phi }}_{0})$ converge weakly to a multivariate normal
distribution with mean vector zero and covariance matrix given by the
covariance function
\begin{equation}
K(x,y):={\mathrm{E}}(\varepsilon _{i}^{2}-1)^{2}\mathbb{I}(Y_{i-1}\leq x\wedge
y)=(\kappa _{\varepsilon }-1)G(x\wedge y),\quad x,y\in {\mathbb{R}},
\label{kxx}
\end{equation}%
where
$G$ denotes the (unconditional) distribution function (d.f.)\thinspace of $Y_{0}$,
$\kappa _{\varepsilon }:={\mathrm{E}}\varepsilon _{i}^{4}<\infty $ and
$x\wedge y=\min (x,y)$.
Under $\mathsf{H}_{0}$, $G$ may depend on ${\boldsymbol{\phi }}_{0}$, but we do not exhibit
this dependence.
Then, since the function $\pi (x):=K(x,x)=%
(\kappa _{\varepsilon }-1)G(x)$ is nondecreasing and nonnegative, tightness
of the process ${\mathcal{U}}_{n}(\cdot ,{\boldsymbol{\phi }}_{0})$ follows by e.g. Theorem 15.7 in \cite{Billingsley:68}, and therefore, under
$\mathsf{H}_{0}$, ${\mathcal{U}}_{n}(\cdot ,{\boldsymbol{\phi }}_{0})$ converges weakly to the time-transformed Brownian motion $B\circ \pi $,
in the space $\cD(\R)$ 
equipped with the uniform~metric.

However, since ${\boldsymbol{\phi }}_{0}$ is replaced by $\hat{{\boldsymbol{%
\phi }}}$, the~weak limit of ${\mathcal{U}}_{n}(\cdot ,\hat{{\boldsymbol{%
\phi }}}\,)$ will not be of the form $B\circ \pi $; rather, it depends on~$({%
\boldsymbol{\phi }}_{0},G)$. We derive this result in the next subsection,
where weak convergence of ${\mathcal{U}}_{n}(\cdot ,\hat{{\boldsymbol{\phi
}}}\,)$ is derived for the case where the true value ${\boldsymbol{\phi }}%
_{0}$ lies in the interior of the parameter space.


\subsection{Asymptotics for the original test statistics}

First we introduce some notation to facilitate the presentation of the underlying assumptions for the asymptotic results.
Let $\cA_{\bfphi}(z) = \sum_{i=1}^{p_1} \al_{i} z^i$ and $\cB_{\bfphi}(z) = 1 - \sum_{i=1}^{p_2} \b_{i} z^i$
with $\cA_{\bfphi}(z) = 0$ if $p_1=0$ and $\cB_{\bfphi}(z) = 1$ if $p_2=0$.
Furthermore, let
\benn
A_{0i} =
\left(
  \begin{array}{cccccc}
   \al_{01}\vep_i^2 & \cdots & \al_{0p_1}\vep_i^2 & \b_{01}\vep_i^2 & \cdots & \b_{0p_2}\vep_i^2 \\
     & \bfI_{p_1-1} & \bf0 &  & \bf0 &  \\
    \al_{01} & \cdots & \al_{0p_1} & \b_{01} & \cdots & \b_{0p_2} \\
     & \bf0 &   &   & \bfI_{p_2-1} & \bf0 \\
  \end{array}
\right), \quad i \ge 1,
\eenn
with $\bfI_k$ denoting the $k \times k$ identity matrix.

In order to study the limiting behaviour of $\cU_n(\cdot, \h \bfphi)$ we make the following assumptions
on the process $\{Y_i\}_{ i \in \mathbb{Z}}$ which satisfies~\eqref{con:var:mod}--\eqref{vsi:mod}.

\begin{Acondition}\label{qml:1}
The parameter space $\Phi$ in equation~\eqref{Phi} is a compact subset of $(0, \iny) \times [0,\iny)^{p_1+p_2}$,
and contains a hypercube of the form
$[\om_L, \om_U] \times [0, \ep]^{p_1+p_2}$, for some $\ep > 0$ and $\om_U > \om_L > 0$,
which includes the true parameter~$\bfphi_0$.
\end{Acondition}

\begin{Acondition}\label{qml:2}
The sequence of matrices $\bfA_0 = (A_{01}, A_{02}, \ldots)$ has a strictly negative top
Lyapunov exponent; i.e., $\g (\bfA_0) = \lim_{i \to \iny} i^{-1} \log \|A_{0i}A_{0(i-1)}\ldots A_{01}\| < 0$,
and $\sum_{j=1}^{p_2} \b_j < 1$, $\forall \bfphi \in \Phi$.
\end{Acondition}


\begin{Acondition}\label{qml:4}
$\cA_{\bfphi_0}(1) \neq 0$, $\al_{0p_1} + \b_{0p_2} \neq 0$,
and the polynomials
$\cA_{\bfphi_0}(z)$ and $\cB_{\bfphi_0}(z)$ have no common roots if $p_2 > 0$.
\end{Acondition}

\begin{Acondition}\label{qml:5}
The errors $\vep_i, i \in \mathbb{Z},$ are i.i.d.\,\,with zero mean and unit variance,
$\vep_i^2$ has a non-degenerate distribution, and
$\mE |\vep_i|^{4+d}  < \iny$ for some $d > 0$.
\end{Acondition}

The condition $\g (\bfA_0) < 0$ in \ref{qml:2} ensures the existence of a unique 
strictly stationary solution  $\{Y_i\}_{ i \in \mathbb{Z}}$ to Model~\eqref{con:var:mod}--\eqref{vsi:mod}; see, e.g. \cite{Bougerol:Picard:92}.
Note that, in~\ref{qml:2}, the strict stationarity condition $\g (\bfA_0) < 0$ is imposed only on the true value $\bfphi_0$,
but for $\bfphi \neq \bfphi_0$ we only impose the weaker restriction $\sum_{j=1}^{p_2} \b_j < 1$. 
In Assumption~\ref{qml:4}, the condition $\cA_{\bfphi_0}(1) \neq 0$  ensures that all the $\al_{0i}$ are not zero when $p_1 \neq 0$,
and hence we do not allow the strictly stationary solution of~\eqref{con:var:mod}--\eqref{vsi:mod} to be a strong white noise process. 
This in turn allows us to avoid certain identifiability issues when estimating the GARCH parameters with $p_2 \neq 0$ (see \citealp{Francq:2010}).
Note that, in the ARCH case (i.e.\,\,when $p_2 =0$), the Assumption~\ref{qml:4} is not required.
In the general GARCH case when $p_2 > 0$, the Assumption~\ref{qml:4} allows for an overidentification of either the order of the ARCH parameters $p_1$
or the order of the GARCH parameters $p_2$, but not both.
The condition $\mE |\vep_i|^{4+d}  < \iny$ in Assumption~\ref{qml:5} is only required for
the existence of the variance of the score vector $\partial \ell_i(\bfphi_0)/\partial \bfphi$; this is necessary for establishing the limiting distribution of the QMLE.
Note that we do not assume that the true parameter $\bfphi_0$ is in the interior of $\Phi$.
Thus, the assumptions do not exclude the cases where some 
$\al_i$ or $\b_j$ are zero.
Assumptions similar to~\ref{qml:1}--\ref{qml:5} have previously been discussed in the literature for establishing asymptotic properties of the QMLE; see, e.g., \cite{Francq:2010} and \cite{Cavaliere:19}.

Let
\benn
J(y,\bfphi) := \mE[\tau_1(\bfphi)\mathbb{I}(Y_{0} \le y)], \quad \tau_i(\bfphi)
:= \frac{(\partial/\partial \bfphi) {\it h}_i(\bfphi)}{{\it h}_i(\bfphi)}, \quad i \in \Z, \, \bfphi \in \Phi.
\eenn
The next lemma provides an asymptotic uniform expansion for $\cU_n(y,\h \bfphi)$.
We make use of this expansion in the proof of establishing the weak convergence of $\cU_n(\cdot,\h \bfphi)$.

\begin{lemma}\label{thm:1}
Suppose that
Assumptions \ref{qml:1} 
and \ref{qml:5} hold.
Then, uniformly in~$y \in \R$,
\ben\label{hU-U}
\cU_n(y,\h \bfphi) =\cU_n(y,\bfphi_0)  - n^{1/2}(\h \bfphi - \bfphi_0)^\prime J(y,\bfphi_0) + o_p(1).
\een
\end{lemma}
Unlike the process $\cU_n(y,\bfphi_0)$, the estimated process $\cU_n(y,\h \bfphi)$ does not converge~weakly to a time transformed Brownian motion,
because the term $n^{1/2}(\h \bfphi-\bfphi_0)^\prime J(y,\bfphi_0)$ in~\eqref{hU-U}, is
of order $O_p(1)$ and hence is not asymptotically negligible.
In fact, if Assumptions~\ref{qml:1}--\ref{qml:4} 
are satisfied, then $\h \bfphi$ converges to $\bfphi_0$ almost surely (a.s.),
and additionally, if Assumption~\ref{qml:5} also holds and $\bfphi_0$ is an interior point in $\Phi$, then
$\h \bfphi$ is asymptotically linear and satisfies
\ben\label{th:expan}
n^{1/2}(\h \bfphi-\bfphi_0) = - \Si_n^{-1}(\bfphi_0) n^{-1/2}\s (1-\vep_i^2)\tau_i(\bfphi_0) + o_p(1),
\een
where
\benn
\Si_n(\bfphi) := n^{-1}\s \tau_i(\bfphi) \tau_i(\bfphi)^\prime, \quad \tau_i(\bfphi)
:= \frac{(\partial/\partial \bfphi) {\it h}_i(\bfphi)}{{\it h}_i(\bfphi)}, \quad \bfphi \in \Phi;
\eenn
see, for example, \cite{Berkes:etal:03}.

By using Lemma~\ref{thm:1} and~\eqref{th:expan}, when $\bfphi_0$ is an interior point in $\Phi$,
one can show that $\cU_n(\cdot,\h \bfphi)$ converges weakly to a centred Gaussian process. 
This result is stated in the next~theorem.

\begin{theorem}\label{thm:2}
Suppose
that \ref{qml:1}--\ref{qml:5} are satisfied with $\bfphi_0$ being an interior point in $\Phi$.
Let
\benn
M_i(\bfphi) := - \Si^{-1}(\bfphi) (1-\vep_i^2)\tau_i(\bfphi), \quad \Si(\bfphi) := \mE \{\tau_1(\bfphi) \tau_1(\bfphi)^\prime\}, \quad \bfphi \in \Phi, \ i \in \Z.
\eenn
Then, 
the process
$\cU_n(\cdot,\h \bfphi)$ converges weakly to $\cU_0$ in $\cD(\R)$,
 where $\cU_0$ is a centred Gaussian process with covariance~kernel
\benrr
\text{\em Cov} \{\cU_0(x), \cU_0(y)\} &=& K(x,y) + J^\prime(x, \bfphi_0) \mE[M_1(\bfphi_0)M_1^\prime(\bfphi_0)] J^\prime(y, \bfphi_0)\\
&& - J^\prime (x, \bfphi_0)\mE[(\vep_1^2-1)M_1(\bfphi_0)\mathbb{I}(Y_0 \le y)]\\
&& - J^\prime (y, \bfphi_0)\mE[(\vep_1^2-1)M_1(\bfphi_0)\mathbb{I}(Y_0 \le x)],
\eenrr
where $K(x,y)$ is as in~\eqref{kxx}.
\end{theorem}
In view of Theorem~\ref{thm:2}, the limiting distributions of KS and CvM statistics defined in~\eqref{eq-T1} depend on the unknown $(\bfphi_0, G)$ in a non-trivial way,
despite the fact the true parameter is in the interior of $\Phi$.
Consequently, it does not appear that it would be possible to find a transformation that would lead to an asymptotically distribution free test, for example as in
\cite{Bai:03,Koul:Indee:12,Indee:Hira:15,Escanciano:etal:18}.
Hence, we proceed by considering bootstrap implementations of the tests.

\subsection{Bootstrap implementation}\label{sec:boot:test}  

In this section, we propose a bootstrap procedure for computing the critical values for the KS and CvM statistics in~\eqref{eq-T1}.
We perform the resampling scheme under the null hypothesis and derive the asymptotic
properties of the bootstrap statistics, irrespective of whether or not the data
generating process satisfies the null hypothesis. To this end, we initially
standardize the residuals
$
\widehat{\varepsilon}_i  :=
Y_i/\{{\it h}_i(\hat{\bfphi})\}^{1/2},  \ i =1, \ldots, n,
$
as
\begin{equation}\label{residuals:scaled}
\check \vep_i := \Big\{n^{-1} \sum_{t=1}^n\bar{\varepsilon}_t^2\Big\}^{-1/2}\bar{\varepsilon}_i, \quad
\bar{\varepsilon}_i := \wh \vep_i - n^{-1} \sum_{t=1}^n\wh \vep_t ,  \quad i = 1, \ldots, n,
\end{equation}
and define the associated empirical distribution function of $\{\check{\varepsilon}_1,\ldots, \check{\varepsilon}_n\}$ as
\ben\label{checkF}
\check {F}_n(x) := n^{-1} \s \mathbb{I}(\check{\varepsilon}_i \leq x), \quad x \in \R.
\een
By construction, $\int_{\mathbb{R}}u\check{F}_{n}(u)du=0$
and $\int_{\mathbb{R}}u^{2}\check{F}_{n}(u)du=1$, hence a random variable
with distribution function $\check{F}_{n}$ has zero mean and unit variance,
therefore matching the first and second order moments of the error
distribution $F_{0}$. From Lemma~\ref{lem:hf} in Appendix~\ref{appendixA} we obtain that $\check{F}_{n}$
converges to $F_{0}$ with probability one under the null hypothesis.

We next outline the bootstrap algorithm.

\subsubsection*{Bootstrap algorithm 1}


%

\noindent {\emph{Step 1:}} Compute $\{\h \bfphi, T_j\}$  on the original sample $\{Y_1, \ldots, Y_n\}$,
where $T_j$ is the test statistic defined in \eqref{eq-T1}~($j=1, 2$);
\newline \noindent { \emph{Step 2:}} Compute $\check{\varepsilon}_{i},\,i=1,\ldots
,n $ as in~\eqref{residuals:scaled} and draw a random sample (with
replacement) of size $n$, say $\{\varepsilon _{1}^{\ast },\ldots
,\varepsilon _{n}^{\ast }\}$, independent of the original data, from the
empirical distribution function $\check F_n(\cdot)$ in~\eqref{checkF};  
%
\newline \noindent { \emph{Step 3:}}
Generate the bootstrap sample $\{{Y}_{1}^{*}, \ldots, Y_n^{*}\}$ with bootstrap true values $(\h \bfphi, \check F_n)$ by
\benn
Y_i^* = \{{\it h}_i^*(\h \bfphi)\}^{1/2}\vep_i^*, \quad {\it h}_i^*(\h \bfphi) = \h \om + \sum_{j=1}^{p_1} \h \al_{j} (Y_{i-j}^*)^2
+ \sum_{k=1}^{p_2} \h \b_{k} {\it h}_{i-k}^*(\h \bfphi), \quad i \ge 1
\eenn
initialized with
$(Y_0^*, \ldots, Y_{1-q}^*, {\it h}_0^{*}(\h \bfphi), \ldots, {\it h}_{1-p}^{*}(\h \bfphi))^\prime = \varsigma_0$, where 
$\varsigma _{0}$ is an arbitrarily chosen vector (e.g. $Y_{t}^*=0$ and ${\it h}_{t}^*=0$, all $t\leq 0$);
\newline { \emph{Step 4:}} Using $\{Y_1^{*}, \ldots, Y_n^{*}\},$
compute $\hat{\bfphi}^{*}$, the bootstrap analog of $\h \bfphi$;
\newline { \emph{Step 5:}} Compute the bootstrap test statistic $T_j^{*}$ as
\ben\label{boot:stats}
T_1^* = \mbox{KS}^*  =\sup_{y}\big|\cU_n^*(y, \h \bfphi^*)\big|, \quad
T_2^* = \mbox{CvM}^*  = \int \left\{\cU_n^*(y, \h \bfphi^*\,)\right\}^2 dG_n^*(y),
\een
where $G_n^*(y)$ and $\cU_n^*(y,\bfphi)$ are the bootstrap analogs of $G_n(y)$ and $\cU_n(y,\bfphi)$, respectively.
%
The bootstrap p-value is then defined as
\begin{equation}
p_{n}^{\ast }:=P_{n}^{\ast }(T_{j}^{\ast }\geq T_{j})  \label{eq bs p val}
\end{equation}%
where $P_{n}^{\ast }$ denotes the probability measure induced by the
bootstrap (i.e., conditional on the original data). The bootstrap test
corresponds to the decision rule:
\begin{equation}
\text{Reject $\mathsf{H}_{0}$ at the nominal level $\alpha $ if the
estimated p-value }p\text{$_{n}^{\ast }$ is less than $\alpha $.}
\label{eq:boot:tst}
\end{equation}
As is standard, $p_{n}^{\ast }$ of (\ref{eq bs p val}) is unknown. It can be
approximated with arbitrary accuracy by repeating steps 2--5 a large number
of times, say $B$, and then setting $p_{n}^{\ast (B)}$ to be the fraction of times
$T_{j}^{\ast }$ exceeds $T_{j}$.

The above bootstrap algorithm is designed to mimic the null data generating process
by replacing the unknown $(\bfphi_0, F_0)$ by the estimators $(\h \bfphi, \check F_n)$.
Therefore, to establish the validity of the bootstrap test in~\eqref{eq:boot:tst},
we need to generalize the regularity assumptions of Lemma~\ref{thm:1} and Theorem~\ref{thm:2} allowing an
arbitrary true value $(\bfphi, F)$ in a neighbourhood of~$(\bfphi_0, F_0)$.
The required conditions are introduced as Assumptions~\ref{A:1}--\ref{A:power} in Appendix~\ref{appendixA}.
Theorem~\ref{thm:3} 
below establishes the asymptotic validity 
of the bootstrap test~\eqref{eq:boot:tst} under \ref{A:1:c}--\ref{A:power}.

In the bootstrap setup, we define $G_n^*(y) := n^{-1}\s \mathbb{I}(Y_{i-1}^*\le y),$ $y \in \R$.
Similarly, the bootstrap analogue of the marked empirical process $\cU_n(y,\bfphi)$ in~\eqref{eq1:0} is defined~by
\ben\label{eq1:0:boot}
\mathcal{U}_n^*(y,\bfphi)
:= n^{-1/2} \sum_{i=1}^n \left\{\frac{(Y_i^*)^2}{{\it h}_i^*(\bfphi)} - 1\right\} \mathbb{I}(Y_{i-1}^* \leq y),
\quad y \in \R, \, \bfphi \in \Phi.
\een
Let $O_{p}^{\ast }(1)$, in probability, $o_{p}^{\ast }(1)$, in probability, and $\mE^{\ast }$
denote the usual stochastic orders of magnitude and expectation,
respectively, with respect to $P_{n}^{\ast }$ defined above. We denote
convergence in distribution of bootstrap statistics as `$\overset{d^{\ast }}{%
\longrightarrow }$'. That is, `$T_{j}^{\ast }\overset{d^{\ast }}{\rightarrow
}g_{j}({\mathcal{U}}_{0})$ \textit{in probability}' means that $%
P_{n}^{\ast }(T_{j}^{\ast }\leq \cdot)\overset{p}{\rightarrow }P\{g%
_{j}({\mathcal{U}}_{0})\leq \cdot\},$ at every continuity point of~$P\{%
g_{j}({\mathcal{U}}_{0})\leq \cdot\}$.

Next theorem establishes the asymptotic validity of the bootstrap test~\eqref{eq:boot:tst} under $\mathsf{H}_0$. 

\begin{theorem}\label{thm:3}
Suppose that Assumptions~\ref{qml:1}--\ref{qml:5} and $\mathsf{H}_0$ are satisfied 
and $\bfphi_0$ is an interior point in $\Phi$. Additionally, assume that Assumptions~\ref{A:1:c}--\ref{A:3:c}
hold with $(\bfphi_0^*,F_0^*) = (\bfphi_0,F_0)$.
Let $\cU_0$ be the limit process appearing in Theorem~\ref{thm:2}.
Then, conditional on $\{Y_1, \ldots, Y_n\}$,
\begin{enumerate}
\item $\cU_n^*(\cdot,\h \bfphi^*)$ converges weakly to $\cU_0$, in probability.
\item $g\{\cU_n^*(\cdot,\h \bfphi^*)\} \stackrel{d^*}{\rightarrow} g\{\cU_0\}$, in probability, %
    for any continuous function $g: \cD(\R) \to \R$.
\item  There exists a continuous functional $g_j: \cD(\R) \to \R$ such that
$T_j^{*} = g_{j}\{\cU_n^*(\cdot,\h \bfphi^*)\} +o_{p}^{\ast }(1)$,
in probability ($j = 1, 2$).
\end{enumerate}
 \end{theorem}
In view of Theorem~\ref{thm:3}, the bootstrap test (\ref{eq:boot:tst}) based on $T_j$ is asymptotically valid under~$\mathsf{H}_0$ ($j = 1, 2$).
The next theorem shows that the bootstrap tests are consistent  under~$\mathsf{H}_1$.

First we need to introduce some notation. Let $(\bfphi_0^*,F_0^*)$ be the probability limit of $(\h \bfphi, \check F_n)$,
such that $\h \bfphi \overset{p}{\rightarrow } \bfphi_0^*$ and $d_2(\check F_n,F_0^*) \overset{p}{\rightarrow } 0$ as $n \to \iny$, where
$d_2(F_X, F_Y)$ is the Mallows metric for the distance between two probability distributions
$F_X$ and $F_Y$ (see also Lemma~\ref{lem:hf} in Appendix~\ref{appendixA}).
Clearly,
under the null hypothesis $\mathsf{H}_0$, we have that $(\bfphi_0^*,F_0^*) = (\bfphi_0,F_0)$.

\begin{theorem}\label{thm:power}
Suppose that $\mathsf{H}_1$ holds. Assume that $(\h \bfphi, \check F_n)$ converges in probability to
$(\bfphi_0^*,F_0^*)$,
the pseudo-true value under $\mathsf{H}_1$.
Additionally, assume that Assumptions~\ref{A:1}--\ref{A:power}
hold, $n^{1/2}(\h \bfphi - \bfphi_0^*) = O_p(1)$,
and~\ref{qml:7} holds if some components of $\bfphi_0^*$ are zero.
Then, conditional on $\{Y_1, \ldots, Y_n\}$,
the bootstrap test (\ref{eq:boot:tst}) based on $T_j$ has asymptotic power~1 
($j = 1, 2$).
\end{theorem}

In view of Theorem~\ref{thm:power}, our tests have asymptotic power against a given alternative
as long as
Assumptions~\ref{A:1}--\ref{A:power} hold,
\ref{qml:7} holds if some components of $\bfphi_0^*$ are zero, and $n^{1/2}(\h \bfphi - \bfphi_0^*) = O_p(1)$.
Assumption~\ref{A:1} 
introduces some regularity conditions in order to ensure the stationarity of the bootstrap data generating process under $\mathsf{H}_1$.
Assumptions~\ref{A:1:c}--\ref{A:3:c} are the same as in Theorem~\ref{thm:3} except that now $(\bfphi_0^*,F_0^*) \neq (\bfphi_0,F_0)$, and
Assumption~\ref{A:power} claims that there exists a $y \in \R$ such that
$\mE[\{{\it h}_1/{\it h}_1(\bfphi_0^*) - 1\}\mathbb{I}(Y_0 \le y)] \neq 0$, 
where ${\it h}_i = \mE(Y_i^2 \mid \cH_{i-1})$, $i \in \Z$.
Since ${\it h}_i$ is not of the form ${\it h}_i(\bfphi)$ under~$\mathsf{H}_1$ and $(\bfphi_0^*,F_0^*)$ is the pseudo-true value, the requirement
$\mE[\{{\it h}_1/{\it h}_1(\bfphi_0^*) - 1\}\mathbb{I}(Y_0 \le y)] \neq 0$ is not very restrictive under $\mathsf{H}_1$.
However, in finite samples, the power of the tests can be sensitive to the form of the discrepancy between ${\it h}_i$ and  ${\it h}_i(\h \bfphi)$.
More precisely, if ${\it h}_i(\h \bfphi)$ is significantly different from ${\it h}_i$
such that the magnitude of the process $n^{-1/2}\s \{Y_i^2/{\it h}_i(\h\bfphi) - 1\}\mathbb{I}(Y_{i-1} \le y)$ is `large' for some $y$,
then the KS and CvM functionals of $n^{-1/2}\s \{Y_i^2/{\it h}_i(\h\bfphi) - 1\}\mathbb{I}(Y_{i-1} \le y)$ are likely to be significantly large
compared to realizations from the empirical distributions of KS$^*$ and CvM$^*$, respectively,
leading to finite sample power of the bootstrap tests.
Importantly, if the true conditional variance ${\it h}_i$ is non-linear while the null parametric form ${\it h}_i(\bfphi)$ is linear,
then our tests are likely to have better finite sample power compared to the case where ${\it h}_i$ and ${\it h}_i(\bfphi)$ are both linear
and the misspecification is only in terms of some missing lags, because in the latter case the
KS and~CvM functionals of $n^{-1/2}\s \{Y_i^2/{\it h}_i(\h\bfphi) - 1\}\mathbb{I}(Y_{i-1} \le y)$ are likely to be smaller compared to the former.

For the validity of our bootstrap tests we have so far required the true value ${%
\boldsymbol{\phi }}_{0}$ to be an interior point of $\Phi $ under $\mathsf{H}_0$. It is of interest
to see whether the bootstrap implementation of $T_{j}(j=1,2)$ can be
modified to obtain a consistent bootstrap test for the case ${\boldsymbol{%
\phi }}_{0}$ lies on the boundary of $\Phi $ under $\mathsf{H}_0$. We consider this in the next
section.

\section{Inference when the true value is on the boundary}\label{sec:4}

Heuristic arguments suggest that $T_1$ and $T_2$ in~\eqref{eq-T1} could serve as possible test statistics for testing $\mathsf{H}_0$ against $\mathsf{H}_1$
regardless of whether 
$\bfphi_0$ lies in the interior or on the boundary of the parameter~space.
In fact, from Lemma~\ref{thm:1}, under assumptions \ref{qml:1} 
and \ref{qml:5}, 
we have
\ben\label{hU-U2}
\cU_n(y,\h \bfphi) =\cU_n(y,\bfphi_0)  - n^{1/2}(\h \bfphi - \bfphi_0)^\prime J(y,\bfphi_0) + o_p(1),
\een
uniformly in~$y \in \R$, irrespective of whether $\bfphi_0$ is in the interior or on the boundary of~$\Phi$,
with $\cU_n(\cdot,\bfphi_0)$ converging weakly to a time transformed Brownian motion.
Therefore, the weak limit of $\cU_n(\cdot, \h \bfphi\,)$, and hence the limiting distributions of $T_1$ and $T_2$,
depend on the asymptotic behaviour of $n^{1/2}(\h \bfphi - \bfphi_0)^\prime J(\cdot,\bfphi_0)$.
%
Hence, to establish the limiting distributions of the test statistics  it is essential to study the large sample properties of $\h \bfphi$ when $\bfphi_0$ lies on the boundary of the parameter space.
Several important results on this have already been obtained by~\cite{Andrews:01} and~\cite{Francq:Zakoian:07}. 
For the ease of reference, in the next subsection,
we summarize some of these results in the notation used in this~paper.

\subsection{Limiting distributions of the estimators}\label{sec:m:res}

In this section, we summarize several technical results regarding the asymptotic behaviour of the
QMLE $\h \bfphi$ in~\eqref{lth} when some components of $\bfphi_0$ are allowed to be zero, and hence $\bfphi_0$ could be on the boundary of $\Phi$.
First, we introduce the following additional regularity condition.

\begin{Acondition}\label{qml:6}
$b_j(\bfphi_0) > 0$ for all $j \ge 0$, where ${\it h}_i(\bfphi_0) = \sum_{j=1}^\iny b_j(\bfphi_0) Y_{i-j}^2$.
\end{Acondition}

Condition~\ref{qml:6} is equivalent to assuming that the ARCH coefficients being nonzero up to the order of the first GARCH coefficient 
that is zero.
As shown in \cite{Berkes:etal:03}, a recursive formula may be obtained to compute $b_j(\bfphi)$ for any given $j$.
Further, we have that $b_j(\bfphi) \to 0$ exponentially fast as $j \to \iny$, uniformly in $\bfphi \in \Phi$.
This means that there exists some $0 < \nu < 1$ such that $\nu^{-j} \sup_{\bfphi \in \Phi} b_j(\bfphi) \rightarrow 0$ as $j \to \iny$.

Since the parameter $\bfphi_0$ is allowed to contain zero components, by the assumption that~$\Phi$ contains a hypercube (see \ref{qml:1}),
the space $n^{1/2}(\Phi - \bfphi_0)$ increases to
the convex cone
\benn
\Lambda = \Lambda(\bfphi_0) = \Lambda_1 \times \Lambda_2 \times \cdots \times \Lambda_{p_1+p_2+1},
\eenn
where $\Lambda_1 = \R$, and for each $i = 2, \ldots, p_1+p_2+1$,
denoting $\bfphi_0 = (\bfphi_{01}, \ldots, \bfphi_{0(1+p_1+p_2)})^\prime$,
$\Lambda_i = \R$ if $\bfphi_{0i} \neq 0$ and $\Lambda_i = [0,\iny)$ if $\bfphi_{0i} = 0$.
Next lemma shows that, under \ref{qml:1}--\ref{qml:6},
the asymptotic distribution~of $n^{1/2}(\h \bfphi - \bfphi_0)$ can be represented as the projection of a normal
vector distribution onto~$\Lambda$;
for further details on the nature of this projection,
see Section~4 in \cite{Francq:Zakoian:07}.

\begin{lemma}\label{thm:1g}
Suppose that Assumptions~\ref{qml:1}-\ref{qml:4} are satisfied.
Then, 
$\h \bfphi \stackrel{a.s.}{\rightarrow} \bfphi_0$, as $n \to\iny$.
Additionally, assume that Assumptions~\ref{qml:5} and \ref{qml:6}
are also satisfied.
Then,
\benn
n^{1/2}(\h \bfphi - \bfphi_0)  \stackrel{d}{\rightarrow} \la^\Lambda := \arg\inf_{\la \in \Lambda} (\la - Z)^\prime \Si(\bfphi_0) (\la - Z),
\eenn
where
$Z \sim \cN (0, (\kappa_{\vep}-1)\Si^{-1}(\bfphi_0)), \ \Si(\bfphi) := \mE \{\tau_1(\bfphi) \tau_1(\bfphi)^\prime\}, \ \bfphi \in \Phi.$
\end{lemma}
The proof of Lemma~\ref{thm:1g} follows from \cite{Francq:Zakoian:07}.
If $\bfphi_0$ is an interior point, 
then $\Lambda = \R^{p_1+p_2+1}$ and $\la^\Lambda = Z \sim \cN (0, (\kappa_{\vep}-1)\Si^{-1}(\bfphi_0))$,
which~is the same as the classical case 
(e.g., see \citealp{Berkes:Horvath:2004})
as we also considered in the previous section.

%

\subsection{Inconsistency of the standard bootstrap test with parameters on the boundary}

The bootstrap true parameter value, say $\bfphi_n^*$, plays a crucial role in defining the properties of any bootstrap test.
For the standard bootstrap test in Section~\ref{sec:boot:test} we set $\bfphi_n^*$ equal to $\h \bfphi$. 
In the proof of Theorem~\ref{thm:3},
under Assumptions~\ref{qml:1}--\ref{qml:5} and~\ref{A:1:c}--\ref{A:3:c}, we obtain that the limiting behaviour of $n^{1/2}(\h \bfphi^* - \bfphi_n^*)$,
conditional  on $(Y_1, \ldots, Y_n)$, is the same as that of $n^{1/2}(\h \bfphi - \bfphi_0)$, under~$\mathsf{H}_0$, since $\bfphi_0$ is an interior point and $\bfphi_n^* = \h \bfphi$.
This result plays a key role in the proof of establishing the validity of the bootstrap tests
for the case the true parameter lies in the interior of the parameter space.
Convergence results of this type have also been used in
establishing the asymptotic validity of other bootstrap methods in similar contexts (see \citealp{Hidalgo:Zaffaroni:07,Perera:Silvapulle:14}).

However, in the current setup,
the parameter $\bfphi_0$ is allowed to contain zero components, and hence we require additional conditions to ensure that the bootstrap tests are consistent.
In particular, 
a crucial requirement for 
the validity of the bootstrap tests 
is to have the following rate of consistency for the bootstrap true value~$\bfphi_n^* = (\bfphi_{n1}^*, \ldots, \bfphi_{n(1+p_1+p_2)}^*)^\prime$:
\ben\label{boot:roc}
n^{1/2}(\bfphi_{ni}^* - \bfphi_{0i}) =
\left\{
  \begin{array}{ll}
    o_p(1), & \hbox{if $\bfphi_{0i}=0$} \\
    O_p(1), & \hbox{if $\bfphi_{0i}>0$}
  \end{array}
\right., \quad
i= 1,2, \ldots, 1+p_1+p_2.
\een
This requirement has previously been introduced in \cite{Cavaliere:19} 
for establishing the validity of a bootstrap based inference procedure in a different context. 
In the current setup, the requirement \eqref{boot:roc} ensures that 
the bootstrap method based on $\bfphi_n^*$ replicates the unknown limiting distribution of $T_j$ under the null, while being of order $O^*_p(1)$, in probability, under the
alternative ($j=1,2$), as is established in Theorems~\ref{thm:6} and~\ref{thm:power:2} below.
If we set $\bfphi_n^* = \h \bfphi$, then it only holds that $n^{1/2}(\bfphi_{ni}^* - \bfphi_{0i}) = O_p(1)$ for $i= 1,2, \ldots, 1+p_1+p_2$,
 and hence \eqref{boot:roc} is not~satisfied.
Therefore, the standard bootstrap test outlined in Section~\ref{sec:boot:test} is not consistent when some parameters lie on the boundary of $\Phi$.
Hence, in the next subsection, instead of the standard bootstrap, we propose a new bootstrap method based on using a different mechanism in choosing the bootstrap true values $\bfphi_{ni}^*$, $i= 1,2, \ldots, 1+p_1+p_2$.

\subsection{Consistent bootstrap implementations}\label{sec:boot:new}

In this section we propose a modified bootstrap testing procedure based on
shrinking the parameter estimators in the bootstrap data generation.
%
The main idea is that, instead of using $\hat{\boldsymbol{\phi }}=\hat{%
\boldsymbol{\phi }}_{n}=(\hat{\boldsymbol{\phi }}_{n1},\ldots ,\hat{%
\boldsymbol{\phi }}_{n(1+p_{1}+p_{2})})^{\prime }$ as the true value in the
bootstrap data generation, we make use of a transformed version of $\hat{%
\boldsymbol{\phi }}$, denoted $\hat{\boldsymbol{\phi }}^{\dag }=\hat{%
\boldsymbol{\phi }}_{n}^{\dag }=(\hat{\boldsymbol{\phi }}_{n1}^{\dag
},\ldots ,\hat{\boldsymbol{\phi }}_{n(1+p_{1}+p_{2})}^{\dag })^{\prime }$ defined~by
\ben\label{shrink:th}
\h \bfphi^\dag_{ni} := \h \bfphi_{ni} \mathbb{I}(\h \bfphi_{ni} > c_n) \quad i= 1,2, \ldots, 1+p_1+p_2,
\een
where $c_n$ is a scalar sequence converging to zero at an appropriate rate:
\ben\label{rate:shrink:th}
c_n \to 0, \quad \text{and} \quad n^{1/2}c_n \to \iny \quad \text{as $n \to \iny$.}
\een
This approach has its roots in the Hodges-Le Cam super-efficient type
estimators, see e.g. \cite{Bickel:etal:98}, \cite{Chatterjee:Lahiri:11}
and \cite{Cavaliere:19}.

In view of the parameter restrictions in~\eqref{Phi}, 
denoting $\bfphi_0 = (\bfphi_{01}, \ldots, \bfphi_{0(1+p_1+p_2)})^\prime$, we have that
$\bfphi_{01} = \om_0 > 0$,
$\bfphi_{0i}=\al_{0(i-1)} \ge 0$ $(i=2,\ldots, 1+p_1),$ and
$\bfphi_{0i}=\b_{0(i-1-p_1)} \ge 0 $ $(i=2+p_1, \ldots, 1+p_1+p_2).$
Thus, $\bfphi_{01}$ is always in the interior, and $\bfphi_{0j}$ is on the boundary of the parameter space only if $\bfphi_{0j}=0$
for some $j \in \{2, 3, \ldots, 1+p_1+p_2\}$; i.e.\,\,some ARCH or GARCH coefficient is zero.
Since $\h \bfphi$ is root-$n$ consistent, the proposed shrinkage in terms of the $c_n$ sequence
ensures that $P(\h \bfphi^\dag_{nj} = 0) \to 1$ as $n \to \iny$ whenever
$\plim \h \bfphi_{nj}=0,$ $j \in \{2, 3, \ldots, 1+p_1+p_2\}$,
where `$\plim$' is the probability limit as $n \to \iny$.
Hence, unlike~$\h \bfphi_{nj}$, in large samples, the transformed estimator $\h \bfphi^\dag_{nj}$ lies on the boundary
of the parameter space with large probability whenever $\bfphi_{0j}$ 
is on the boundary; i.e.\,\,$\bfphi_{0j}=0$.
Since $n^{1/2}(\h \bfphi - \bfphi_0)=O_p(1)$ and $c_n$ converges at a rate slower than $n^{-1/2}$,
this ensures that the requirement~\eqref{boot:roc} is satisfied by the parameter $\h \bfphi^\dag_n$ defined by~\eqref{shrink:th}--\eqref{rate:shrink:th}.
Hence, as established in Theorems~\ref{thm:6} and~\ref{thm:power:2} below,
the bootstrap based on $\bfphi_n^*=\h \bfphi^\dag_n$ allows us to replicate the unknown limiting
distributions of $T_1$ and $T_2$ under $\mathsf{H}_0$, while being of order $O_p^*(1)$, in probability, under the~alternative.

We next provide a step-by-step guide of the proposed modified bootstrap~approach.

\subsubsection*{Bootstrap algorithm 2 (shrinking parameter estimators approach)} %

%

\noindent {\emph{Step 1:}} Compute $\{\h \bfphi, T_j\}$  on the original  sample $\{Y_1, \ldots, Y_n\}$;
\newline \noindent { \emph{Step 2:}} Compute $\check{\varepsilon}_{i},\,i=1,\ldots
,n $ as in~\eqref{residuals:scaled} and draw a random sample (with
replacement) of size $n$, say $\{\varepsilon _{1}^{\ast },\ldots
,\varepsilon _{n}^{\ast }\}$, independent of the original data, from the
empirical distribution function $\check F_n(\cdot):= n^{-1} \s \mathbb{I}(\check{\varepsilon}_i \leq \cdot)$;
%
\newline \noindent { \emph{Step 3:}}
Generate the bootstrap sample $\{{Y}_{1}^{*}, \ldots, Y_n^{*}\}$ with bootstrap true values $(\h \bfphi^\dag, \check F_n)$
as%
\benn
Y_i^* = \{{\it h}_i^*(\h \bfphi^\dag)\}^{1/2}\vep_i^*, \quad {\it h}_i^*(\h \bfphi^\dag) = \h \om^\dag + \sum_{j=1}^{p_1} \h \al_{j}^\dag (Y_{i-j}^*)^2
+ \sum_{k=1}^{p_2} \h \b_{k}^\dag {\it h}_{i-k}^*(\h \bfphi^\dag), \quad i \ge 1
\eenn
initialized with
$(Y_0^*, \ldots, Y_{1-p_1}^*, {\it h}_0^{*}(\h \bfphi^\dag), \ldots, {\it h}_{1-p_2}^{*}(\h \bfphi^\dag))^\prime = \varsigma_0$,
where $\varsigma_0$ is 
an arbitrarily chosen vector (e.g., set $Y_{t}^*=0$ and ${\it h}_{t}^*=0$, all $t\leq 0$);
\newline { \emph{Step 4:}} Using
$\{Y_1^{*}, \ldots, Y_n^{*}\},$
compute $\{\hat{\bfphi}^{*}, T_j^{*} \}$ the bootstrap analogs of $\{\h \bfphi,  T_j\}$.
%

\noi The bootstrap test then corresponds to the decision rule:
\begin{equation}\label{eq:boot:tst:2}
\text{Reject $\mathsf{H}_{0}$ at the nominal level $\alpha $ if the
estimated p-value }p\text{$_{n}^{\ast }$ is less than $\alpha $.}
\end{equation}
The bootstrap p-value $p_{n}^{\ast }$ is defined as in~(\ref{eq bs p val}), and it can be
approximated with arbitrary accuracy by repeating steps 2--4 a large number
of times, say $B$, and then setting $p_{n}^{\ast (B) }$ to be the fraction of times
$T_{j}^{\ast }$ exceeds $T_{j}$ ($j=1, 2$).

Note that, the limiting distribution of $n^{1/2}(\h \bfphi^\dag - \bfphi_0)$ is the same as that of $n^{1/2}(\h \bfphi - \bfphi_0)$ whenever $\bfphi_0$ is in the interior of $\Phi$.
Hence, the bootstrap test~\eqref{eq:boot:tst:2} collapses into the bootstrap method outlined in Section~\ref{sec:boot:test} as $n \to \iny$, whenever $\bfphi_0$ is in the interior of $\Phi$.

\subsection{Asymptotic validity} 

In this section we establish the asymptotic validity of the bootstrap based on the shrinking parameter estimators approach
introduced in the previous subsection. 

Note that the bootstrap analogue of the marked empirical process $\cU_n(y,\bfphi)$ for the bootstrap test~\eqref{eq:boot:tst:2} is defined as in~\eqref{eq1:0:boot}, with
\benn\label{eq1:boot:sp}
\mathcal{U}_n^*(y,\bfphi)
:= n^{-1/2} \sum_{i=1}^n \left\{\frac{(Y_i^*)^2}{{\it h}_i^*(\bfphi)} - 1\right\} \mathbb{I}(Y_{i-1}^* \leq y),
\quad y \in \R, \, \bfphi \in \Phi,
\eenn
except that $Y_i^*$ and ${\it h}_i^*(\bfphi)$ are now based on the bootstrap method outlined in Section~\ref{sec:boot:new}.
The next theorem establishes the asymptotic validity of the bootstrap test~\eqref{eq:boot:tst:2}.
First, we introduce the following additional assumption.
Recall that $F_0$ is the c.d.f\,\,of $\vep_i$ in~\eqref{con:var:mod}.

\begin{Acondition}\label{qml:7}
The Assumptions~\ref{qml:1}--\ref{qml:6} continue to hold
when $\zeta_0=(\bfphi_0, F_0)$ is replaced by $\zeta_n=(\bfphi_n, F_n)$, where $\zeta_n \ra (\bfphi_0^*, F_0^*):=\plim (\h \bfphi, \check F_n)$ as $n \to \iny$.
\end{Acondition}

Since Assumptions~\ref{qml:1}--\ref{qml:6} correspond to the original data generating process, the underlying true parameter value $\zeta_0=(\bfphi_0, F_0)$ is fixed.
However, in the bootstrap data generation the true parameter $(\h \bfphi^\dag, \check F_n)$ is not fixed but converges to $(\bfphi_0^*, F_0^*)$ as $n \to \iny$.
Therefore, it is not adequate to assume only~\ref{qml:1}--\ref{qml:6} in order to establish the validity of the bootstrap tests.
Assumption~\ref{qml:7} ensures that \ref{qml:1}--\ref{qml:6} hold for triangular arrays, and hence allows us to extend the arguments in the proof of Lemma~\ref{thm:1g} to a triangular array setup,
which in turn is essential for establishing that the limiting distribution of $n^{1/2}(\h \bfphi^* - \h \bfphi^\dag_n)$,
conditional  on $(Y_1, \ldots, Y_n)$, is the same as that of $n^{1/2}(\h \bfphi - \bfphi_0)$
under $\mathsf{H}_0$, while being of order $O_p^*(1)$, in probability, under $\mathsf{H}_1$.
This result plays a key role in the proofs of the asymptotic validity and consistency of the bootstrap tests obtained in the next two theorems.


\begin{theorem}\label{thm:6}
Suppose that Assumptions~\ref{qml:1}--\ref{qml:7} and \ref{A:1:c}--\ref{A:3:c} hold.
Then, under $\mathsf{H}_0$, the conditional weak limit of $\cU_n^*(\cdot,\h \bfphi^*)$ is the same as that of $\cU_n(\cdot,\h \bfphi)$, in probability,
and hence the bootstrap test~\eqref{eq:boot:tst:2} based on $T_j$ is asymptotically valid ($j = 1, 2$).
\end{theorem}

The next theorem establishes the consistency of the bootstrap test~\eqref{eq:boot:tst:2}.

\begin{theorem}\label{thm:power:2}
Suppose that $\mathsf{H}_1$ holds.
Assume that the estimator $\h \bfphi$
converges in probability to some point in $\Phi$,
and $\check F_n$ in~\eqref{checkF} converges in probability with respect to the Mallows metric.
Suppose that Assumptions~\ref{A:1}--\ref{A:3:c} hold
with $(\bfphi_0^*, F_0^*):=\plim (\h \bfphi, \check F_n)$. 
Additionally, assume that Assumption~\ref{qml:7} holds, $n^{1/2}(\h \bfphi - \bfphi_0^*) = O_p(1)$ and there exists a $y \in \R$, with ${\it h}_i = \mE(Y_i^2 \mid \cH_{i-1})$, $i \in \Z$, such that
$\mE[\{{\it h}_1/{\it h}_1(\bfphi_0^*) - 1\}\mathbb{I}(Y_0 \le y)] \neq 0$.
Then, conditional on $\{Y_1, \ldots, Y_n\}$,
the bootstrap test (\ref{eq:boot:tst:2}) based on $T_j$ has asymptotic power~1~($j = 1, 2$).
 \end{theorem}

Theorem~\ref{thm:6} shows that
the proposed shrinkage in terms of the $c_n$ sequence, or more generally, the requirement~\eqref{boot:roc}
ensures that the bootstrap test statistics $T_1^*$ and $T_2^*$ based on~(\ref{eq:boot:tst:2}) replicate the
unknown limiting distributions of $T_1$ and $T_2$ under the null hypothesis.
Theorem~\ref{thm:power:2} establishes that $T_1^*$ and $T_2^*$ are of order $O_p^*(1)$, in probability, under the~alternative;
that is, the proposed bootstrap method is also consistent even if it is unknown whether any of the nuisance parameters are on the boundary or not.



\section{Numerical Study}\label{Sim-boot-bound}

In this section we carry out a Monte Carlo simulation study to evaluate the
finite sample performance of the KS and CvM tests based on the bootstrap method~\eqref{eq:boot:tst:2} in
Section~\ref{sec:boot:new}.
Our main focus is the case where the true
parameter value ${\boldsymbol{\phi }}_{0}$ of the data generating process
lies on the boundary of the parameter space. For comparison, we also
consider the case where ${\boldsymbol{\phi }}_{0}$ is an interior point.
Several data generating processes under the alternative hypothesis are also considered
in order to investigate the finite sample power properties of the~tests.
Although 
there are several other tests that can be applied for testing the
conditional variance specification in GARCH-type models, 
as mentioned in the introduction,
the theory for their validity
does not hold when the true parameter is on the boundary.
 Hence, in these simulations, we compare the proposed tests with the general purpose Ljung-Box $Q$ test
 which tests the significance of the serial dependence of the squared residuals estimated from the fitted model.
 We denote the Ljung-Box $Q$ test for a lag length $\ell$ by $\text{LBQ}(\ell)$.

\subsection{Design of the simulation study}

Tests are evaluated when the parametric form ${\it h}_i(\bfphi)$~under $\mathsf{H}_0$ is
\benrr
\text{$\mathsf{H}_0^A$ [GARCH(1,1)]:}&& {\it h}_i(\bfphi) = \om + \al Y_{i-1}^2 +\b{\it h}_{i-1}(\bfphi),\\ 
\text{$\mathsf{H}_0^B$ [GARCH(1,2)]:}&& {\it h}_i(\bfphi) = \om + \al Y_{i-1}^2 + \b_1{\it h}_{i-1}(\bfphi) + \b_2{\it h}_{i-2}(\bfphi).
\eenrr
For the error distribution, we consider the standard normal distribution.

For the conditional variance ${\it h}_i$ of the true data generating process [DGP] 
we consider the following 9 cases:

\noindent DGP1 [ARCH(1)]:
${\it h}_i = 0.20 + 0.7 Y_{i-1}^2,$       \hspace*{\fill}{[$\mathsf{H}_0^A$ and $\mathsf{H}_0^B$ both true]}

\noindent DGP2 [GARCH(1,1)]:
${\it h}_i = 0.10 + 0.20 Y_{i-1}^2 + 0.70 {\it h}_{i-1},$    \hspace*{\fill}{[$\mathsf{H}_0^A$ and $\mathsf{H}_0^B$ both true]}

\noindent  DGP3 [GARCH(1,2)]:
${\it h}_i = 0.10 + 0.10 Y_{i-1}^2 +0.15 {\it h}_{i-1} + 0.70 {\it h}_{i-2},$      \hspace*{\fill}{[$\mathsf{H}_0^A$ false and $\mathsf{H}_0^B$ true]}

\noindent DGP4 [GJR-GARCH(1,1)]\footnote{The Glosten-Jagannathan-Runkle GARCH model of~\cite{Glosten:GJR:93}.}:       \hspace*{\fill}{[$\mathsf{H}_0^A$ and $\mathsf{H}_0^B$ both false]}
\newline ${\it h}_i = 0.10 + 0.1 Y_{i-1}^2 +0.5 {\it h}_{i-1} + 0.3 Y_{i-1}^2 \mathbb{I}(Y_{i-1}<0)$,

\noindent  DGP5 [GARCH(2,2)]:
\newline ${\it h}_i = 0.10 + 0.05 Y_{i-1}^2 + 0.6 Y_{i-2}^2 +0.1 {\it h}_{i-1} + 0.2 {\it h}_{i-2},$
\hspace*{\fill}{[$\mathsf{H}_0^A$ and $\mathsf{H}_0^B$ both false]}

\noindent DGP6 [EGARCH(1,1)]:
\newline $\ln {\it h}_i = 0.1 + 0.4 \ln {\it h}_{i-1} + 0.2 ( |\vep_{i-1}| - \mE |\vep_{i-1}| ) - 0.2\vep_{i-1},$
\hspace*{\fill}{[$\mathsf{H}_0^A$ and $\mathsf{H}_0^B$ both false]}

\noindent DGP7 [i.i.d.]:
${\it h}_i = 1,$
 \hspace*{\fill}{[$\mathsf{H}_0^A$ and $\mathsf{H}_0^B$ both true, but Assumption~\ref{qml:4} false]}

\noindent DGP8 [Threshold GARCH(1,1)]:
\newline ${\it h}_i = 0.10 + 0.1 Y_{i-1}^2 +0.5 {\it h}_{i-1} + 0.3 {\it h}_{i-1} \mathbb{I}(Y_{i-1}<0),$
\hspace*{\fill}{[$\mathsf{H}_0^A$ and $\mathsf{H}_0^B$ both false]}

\noindent DGP9 [T-CHARM]\footnote{T-CHARM refers to the conditionally heteroscedastic AR models proposed by \cite{Chan:Tong:14}.}:
 ${\it h}_i = \mathbb{I}(Y_{i-1}\le 0) + 1.2\mathbb{I}(Y_{i-1}>0).$ \hspace*{\fill}{[$\mathsf{H}_0^A$ and $\mathsf{H}_0^B$ both false.]}

DGPs 1, 2, 3, 5 and 7 are linear models. We use these DGPs to evaluate the size and~power properties of the tests,
focusing particularly on the cases where the true parameter lies on the boundary. 
The DGPs 4, 6, 8 and~9 are nonlinear models, and hence the linear models specified under the null hypotheses $\mathsf{H}_0^A$ and $\mathsf{H}_0^B$
are misspecified for each of DGPs 4, 6, 8 and~9. 
We use these four DGPs to evaluate the finite sample power properties of the~tests.

In the next two subsections we consider testing for the two null models $\mathsf{H}_0^A$ [GARCH(1,1)] and $\mathsf{H}_0^B$ [GARCH(1,2)] separately.
The results are based on $2000$ Monte Carlo replications.
For each replication and data generating process, we first compute the QMLE $\h \bfphi$  and compute the test statistics KS, CvM, and LBQ($\ell$), $\ell = 3,5,10,15,20$.
To implement the proposed KS and CvM tests we use the bootstrap method~\eqref{eq:boot:tst:2} outlined in Section~\ref{sec:boot:new} with $c_n=n^{-1/3}/50$,
while adopting the `Warp-Speed' Monte Carlo method of \cite{Giacomini:etal:07} in order to reduce the computational burden. 
The results are presented in Figures~\ref{fig:Hoa:dgp:1:3}--\ref{fig:dgp:tcharm}.
In these figures the results of the LBQ($\ell$) tests are presented for only $\ell = 5, 10$ and $15$;
the patterns of the results for $\ell = 3$ and $\ell = 20$ are similar to those for $\ell = 5, 10$ and $15$ and hence are omitted.

\subsection{Testing for $\mathsf{H}_0^A$ [GARCH(1,1)]}

\begin{figure}[H]
\caption{Empirical rejection rates for testing $\mathsf{H}_0^A$ [GARCH(1,1)]
}
\centering
\includegraphics[width=6.3in]{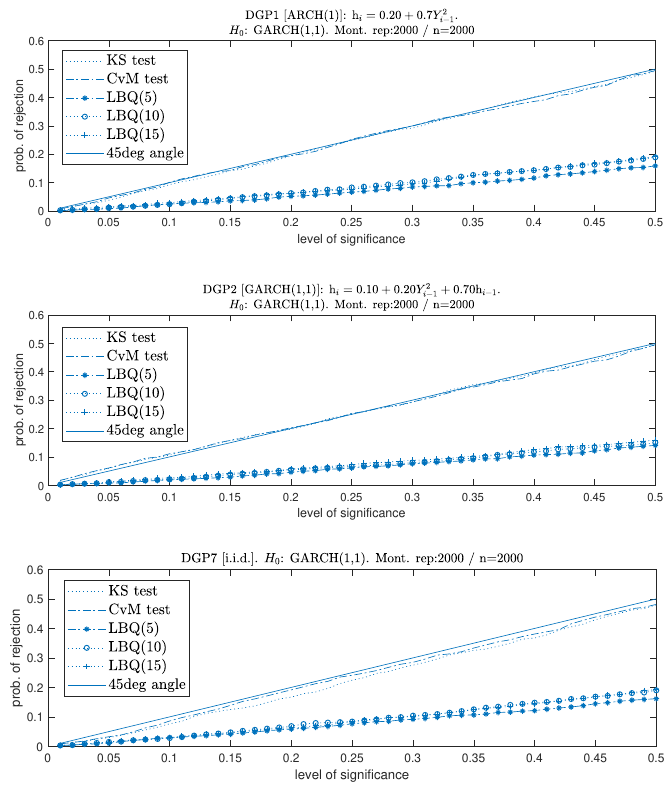}
\label{fig:Hoa:dgp:1:3}
\vspace{-0.1in}
\begin{quote}
{\small
Notes: DGPs 1 and 2 are under $\mathsf{H}_0^A$.
The true parameter for the DGP1 is a boundary point.
For the DGP2, the true parameter is an interior point.
Although the DGP7 [i.i.d.] is a member of the null family $\mathsf{H}_0^A$, with the true value on the boundary,
it does not satisfy all the conditions assumed for establishing the validity of the bootstrap~tests.
}
\end{quote}
\end{figure}

\begin{figure}[H]
\caption{Empirical power of KS and CvM for testing $\mathsf{H}_0^A$ [GARCH(1,1)]
}
\centering
\includegraphics[width=\textwidth]{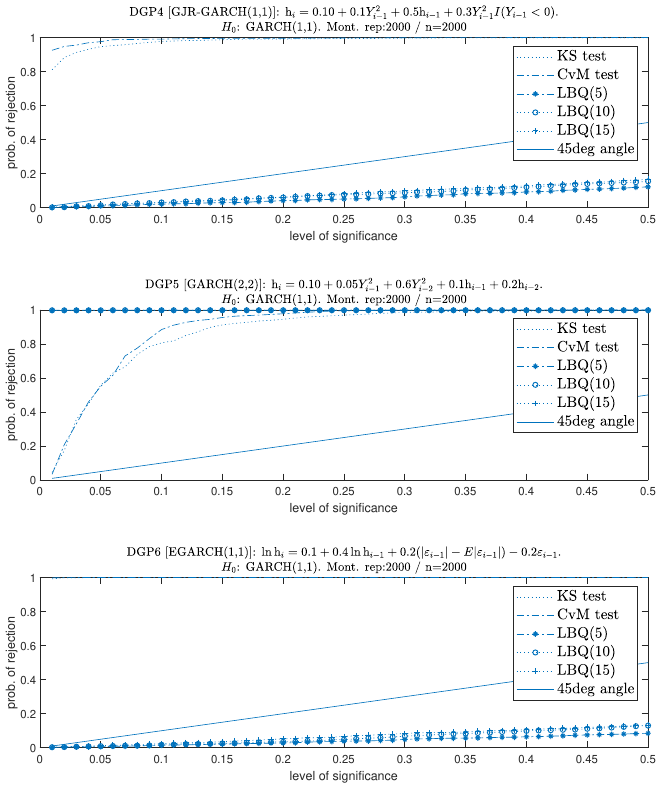}
\label{fig:Hoa:dgp:4:6}
\vspace{-0.1in}
\begin{quote}
{\small
{Notes}: The DGPs 4--6 are considered to evaluate the power of the tests.
}
\end{quote}
\end{figure}

The DGP1 [ARCH(1)] and DGP2 [GARCH(1,1)] are members of the null family under $\mathsf{H}_0^A$.
For the DGP1, the GARCH coefficient is zero, and hence the true parameter lies on the boundary of the parameter space.
For the DGP2, the true parameter is an interior point. 
The DGP7 [i.i.d.] 
is a member of the null family $\mathsf{H}_0^A$ with the true value on the boundary.
However, since there are no ARCH or GARCH coefficients, the DGP 7 does not satisfy Assumption~\ref{qml:4} and hence the theory developed in this paper does not apply in this case.
The DGPs 3, 4, 5, 6, 8 and 9 are considered in order to evaluate the empirical power properties of the tests; these DGPs are part of the alternative hypothesis.
A summary of the results are given in Figures~\ref{fig:Hoa:dgp:1:3}, \ref{fig:Hoa:dgp:4:6}, \ref{fig:dgp:tgarch} and~\ref{fig:dgp:tcharm},
and are discussed in Section~\ref{sim:sum}.

\subsection{Testing for $\mathsf{H}_0^B$ [GARCH(1,2)]}

The DGPs 1, 2 and 3 are members of the null family $\mathsf{H}_0^B$ [GARCH(1,2)].
For each of DGP1 and DGP2, one of ARCH or GARCH coefficients is zero, and hence the true parameter lies on the boundary of the parameter space.
The true parameter of the DGP3 [GARCH(1,2)] is an interior point.
Since there are no ARCH or GARCH coefficients, when testing for $\mathsf{H}_0^B$, the DGP7 does not satisfy Assumption~\ref{qml:4} and hence as in the previous case, the theory developed in this paper does not apply.
The DGPs 4, 5, 6, 8 and 9 continue to be under the alternative hypothesis when testing for GARCH(1,2).

\begin{figure}[H]
\caption{Empirical rejection rates for testing $\mathsf{H}_0^B$ [GARCH(1,2)] when the true parameter lies on the boundary of the parameter space.
}
\centering
\includegraphics[width=\textwidth]{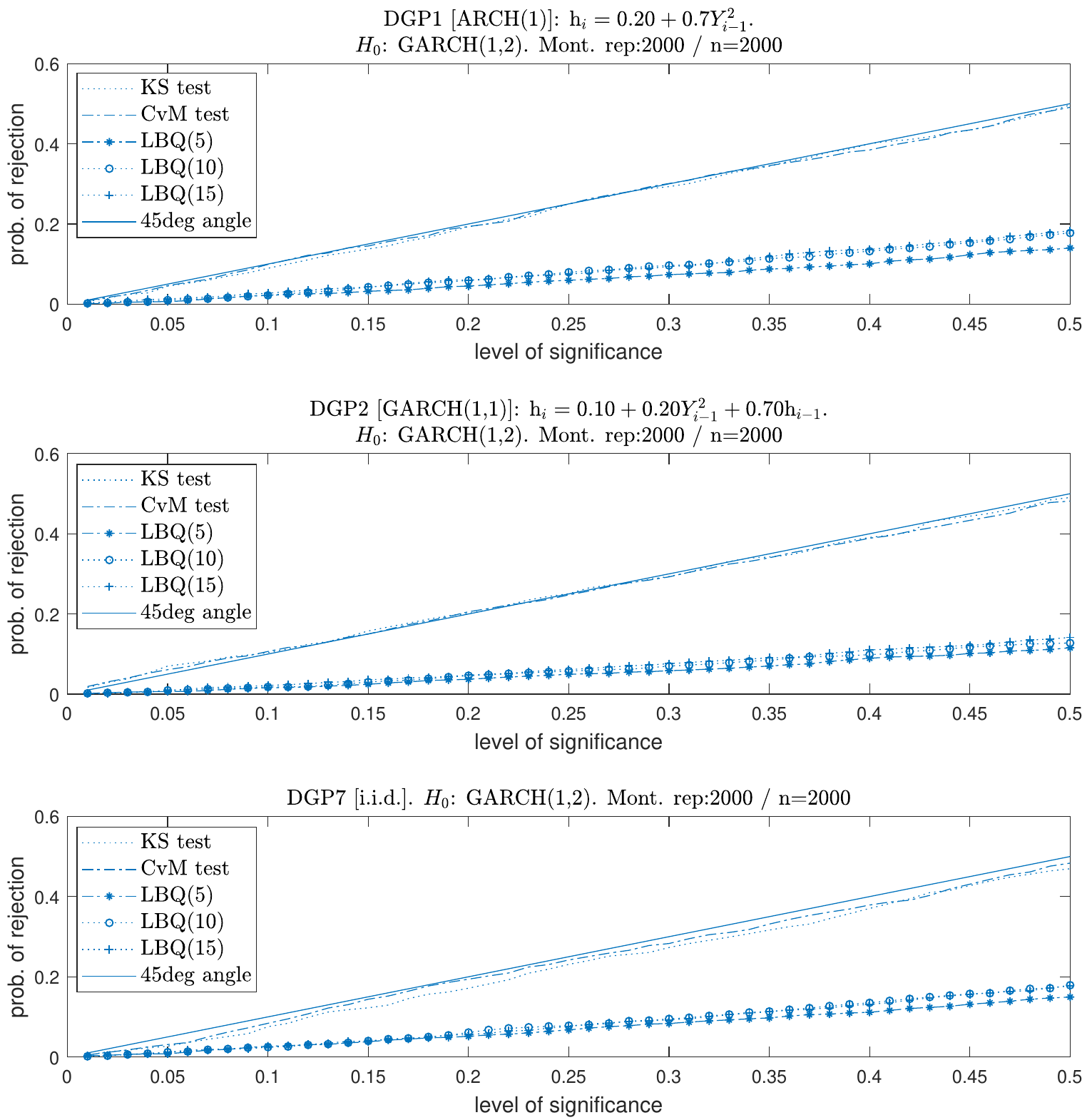}
\label{fig:Hob:dgp:1:3}
\vspace{-0.1in}
\begin{quote}
{\small
{Notes}: For DGPs 1 and 2, the true parameter under $\mathsf{H}_0^B$ lies on the boundary of the parameter space.
}
\end{quote}
\end{figure}

\begin{figure}[H]
\caption{Empirical size 
and power for testing $\mathsf{H}_0^B$ [GARCH(1,2)].
The sample size $n=2000$.
}
\centering
\includegraphics[width=\textwidth]{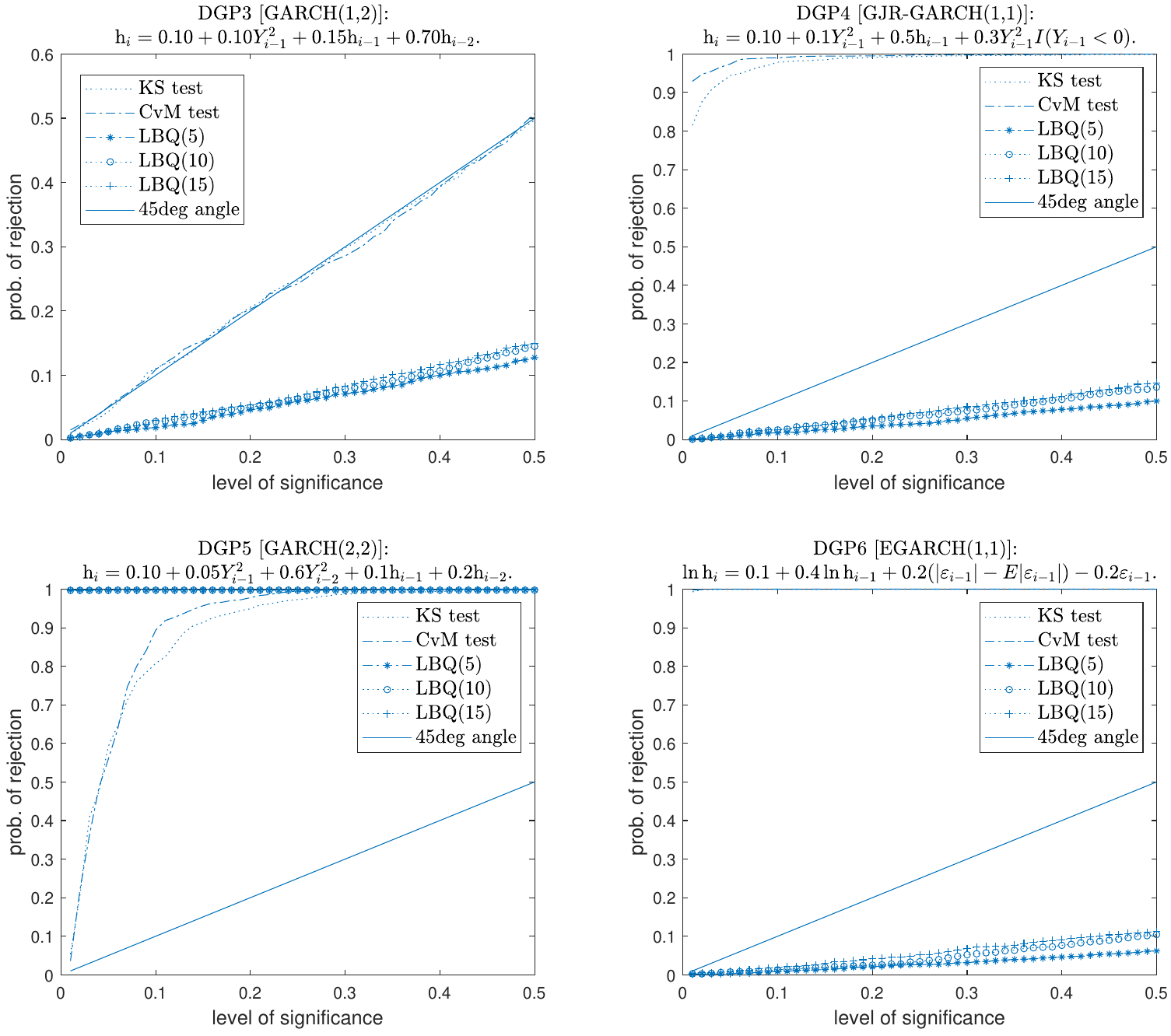}
\label{fig:Hob:dgp:4:6}
\vspace{-0.1in}
\begin{quote}
{\small {Notes}: The DGP3 [GARCH(1,2)] is under $\mathsf{H}_0^B$;
the true parameter of the DGP3 is an interior point of the parameter space.
The DGPs 4, 5, and~6 are part of the alternative.
}
\end{quote}
\end{figure}

\begin{figure}[H]
\caption{Empirical power for testing $\mathsf{H}_0^A$ [GARCH(1,1)] and $\mathsf{H}_0^B$ [GARCH(1,2)]
for the DGP8 [Threshold GARCH(1,1)]: ${\it h}_i = 0.10 + 0.1 Y_{i-1}^2 +0.5 {\it h}_{i-1} + 0.3 {\it h}_{i-1} \mathbb{I}(Y_{i-1}<0)$.
KS \tecb{$\cdots$}, CvM \tecb{-- $\cdot$ -- $\cdot$ --}, LBQ(5) \tecb{-- $\cdot$ $\ast$ $\cdot$ --}, LBQ(10) \tecb{-- $\cdot$ o $\cdot$ --},
LBQ(15) \tecb{-- $\cdot$ $+$ $\cdot$ --}, 45deg \tecb{--}.
}
\centering
\includegraphics[width=\textwidth]{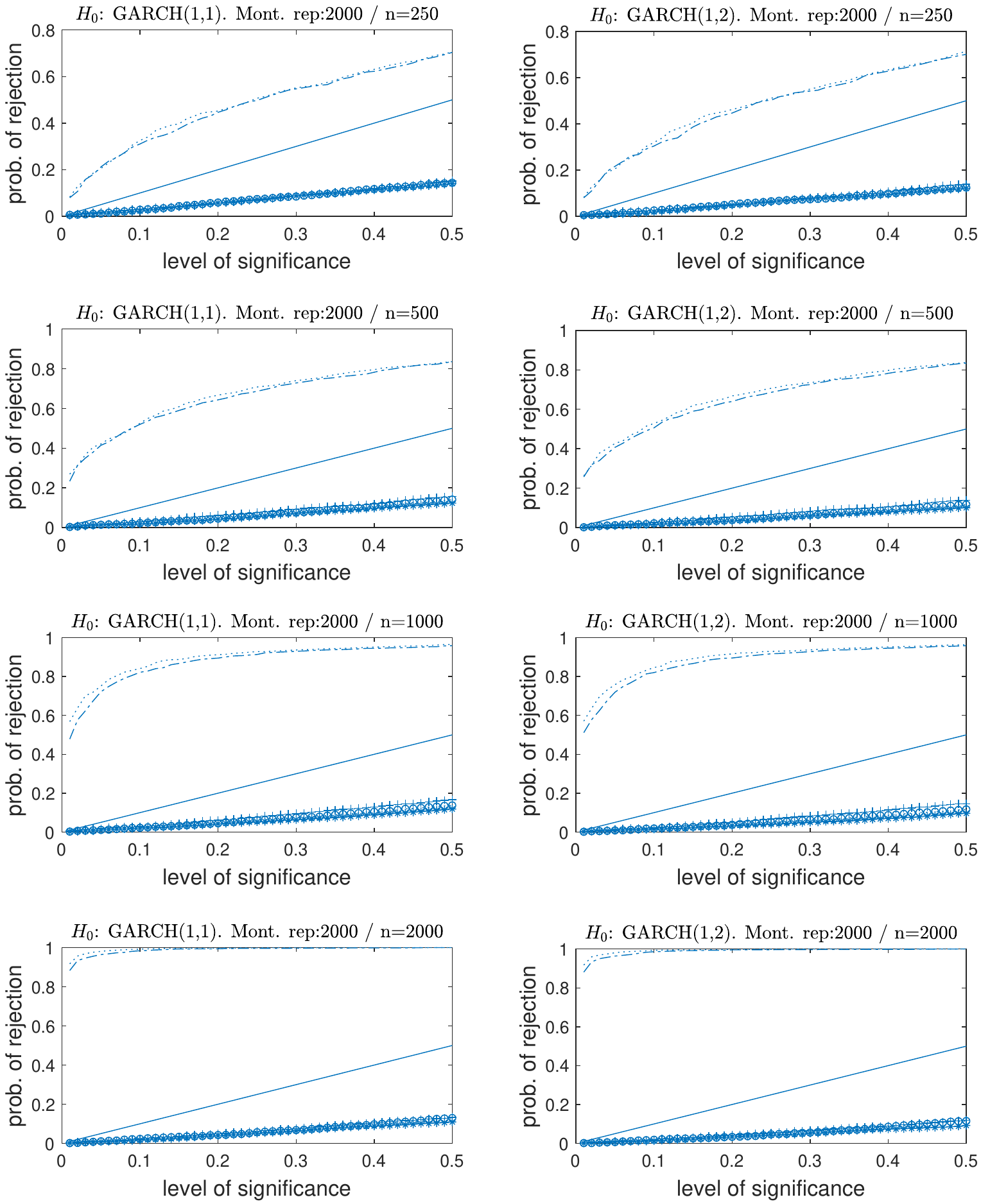}
\label{fig:dgp:tgarch}
\vspace{-0.3in}
\begin{quote}
{\small {Notes}: GARCH(1,1) and GARCH(1,2) are both misspecified for the DGP8. 
}
\end{quote}
\end{figure}

\begin{figure}[H]
\caption{Empirical power for testing $\mathsf{H}_0^A$ [GARCH(1,1)] and $\mathsf{H}_0^B$ [GARCH(1,2)]
for the DGP9 [T-CHARM]: ${\it h}_i = \mathbb{I}(Y_{i-1}\le 0) + 1.2\mathbb{I}(Y_{i-1}>0)$.
KS \tecb{$\cdots$}, CvM \tecb{-- $\cdot$ -- $\cdot$ --}, LBQ(5) \tecb{-- $\cdot$~$\ast$~$\cdot$~--}, LBQ(10) \tecb{-- $\cdot$ o $\cdot$ --},
LBQ(15) \tecb{-- $\cdot$ $+$ $\cdot$ --}, 45deg angle \tecb{--}.
}
\centering
\includegraphics[width=6.3in]{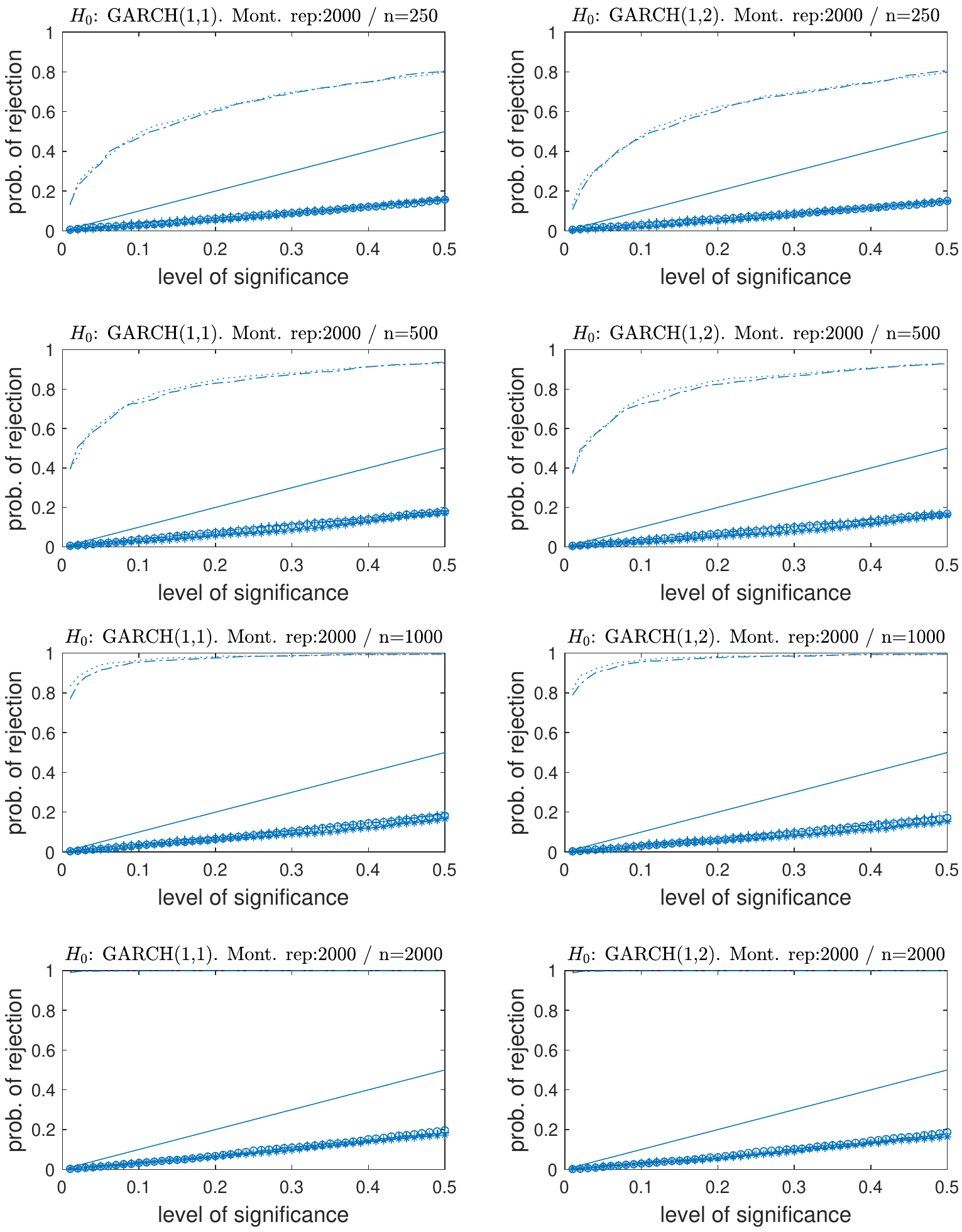}
\label{fig:dgp:tcharm}
\vspace{-0.15in}
\begin{quote}
{\small {Notes}: GARCH(1,1) and GARCH(1,2) are both misspecified for the DGP9.
}
\end{quote}
\end{figure}

The results are given in Figures~\ref{fig:Hob:dgp:1:3}, \ref{fig:Hob:dgp:4:6}, \ref{fig:dgp:tgarch} and~\ref{fig:dgp:tcharm}, and
are discussed in the next subsection.

\subsection{Summary of the results}\label{sim:sum}

The main observations of the simulation results are the following:
\begin{description}
\item[(i) ]
The KS and CvM bootstrap tests proposed in this paper perform consistently well in terms of the
Type-I error rate. 
By contrast the LBQ test does not perform well in terms of finite sample size at any of the lag lengths considered.
In particular, for every instance in which the DGP is under the null hypothesis, the LBQ test is significantly undersized at each of the lag lengths considered.
\item[(ii) ]
No significant difference in performance of the KS and CvM tests can be identified,
in terms of Type-I error rates,
irrespective of whether the true parameters are on the boundary or not.
Thus, the simulation results indicate that the bootstrap method
based on the shrinking parameter estimators approach
outlined in Section~\ref{sec:boot:new} performs well irrespective of whether the true value of $\bfphi$
is on the boundary or in the interior. 
\item[(iii) ]
Both KS and CvM tests exhibit good overall power properties. In particular, they exhibit excellent power gains compared to the LBQ test when the DGP is non-linear.
Although the LBQ test performs better than KS and CvM when the DGP is linear
and the misspecification is in terms of some missing lags,
the power gains are not as significant as in the cases where KS and CvM outperform LBQ.
In particular,
the LBQ test does not exhibit any empirical power against the nonlinear DGPs, DGP4, DGP6, DGP8
and DGP9, %
when testing for both $\mathsf{H}_0^A$ [GARCH(1,1)] and $\mathsf{H}_0^B$ [GARCH(1,2)].
As expected,
the empirical power of KS and CvM tests increase with the significance level~$\al$ and the sample size $n$.
Overall, in terms of empirical power, the CvM test performs
marginally better than the KS test.
\end{description}

%

\section{Empirical illustrations}\label{sec:example}
%
%

To illustrate the bootstrap testing procedure, we briefly discuss two real data examples.

\subsubsection*{Example 1}
We first consider a data example based on the daily log returns of the SPDR exchange-traded fund (ETF) for the S\&P 500
index. This ETF is usually denoted by the tick symbol
SPY. The data spans the period January 3, 2007 to June 30, 2017 and contains 2640 observations.
A shorter version of this data set was previously studied by \cite{tsay2018}, and by using some preliminary diagnostics, they concluded that a GJR-GARCH(1,1) model provides a good fit.
In their empirical analysis, \cite{tsay2018} concluded that the leverage effect of the fitted GJR-GARCH(1,1) model is statistically significant at the 5\% level.
This indicates that if one specifies a GARCH(1,2) model for the conditional variance then that may not provide a good fit for the data.
In order to investigate this, 
in this empirical illustration, we employ the proposed KS and CvM bootstrap tests to
 test the adequacy of the  GARCH(1,2) specification, 
 expecting that the proposed tests would be able to detect a misspecification.
 For comparison the LBQ test considered in the simulations in the previous section is also~considered.

\begin{figure}[H]
\caption{Autocorrelogram of the squared SPY log returns time series (first panel), and the squared residual
correlogram for the fitted GARCH(1,2) model (second panel).
The sample period is from January 3, 2007 to June 30, 2017.
}
\centering
\includegraphics[width=\textwidth]{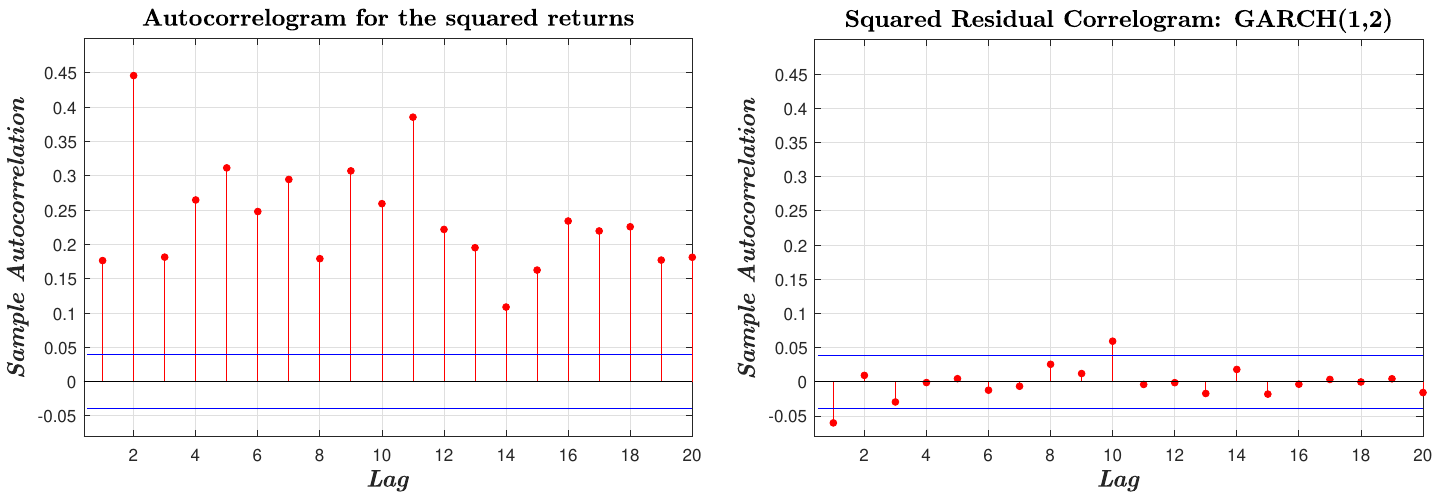}
\label{fig:corr:spy}
\end{figure}

 For the  GARCH(1,2) specification, the $p$-values of the KS and CvM tests are both zero up to 3 decimal places, whereas the $p$-value for the LBQ(20) turns out to be 0.123, and those for LBQ(15) and LBQ(5) are
 0.032 and 0.033, respectively.
 Thus, the KS and CvM tests proposed in this paper clearly reject the GARCH(1,2) specification, but the LBQ(20) fails to reject the GARCH(1,2) specification
 at the 10\% level of significance.
 Note that the Ljung-Box $Q$ test is designed to check the significance of the 
 autocorrelations of the squared residuals at multiple lags jointly.
 Figure~\ref{fig:corr:spy} shows the sample autocorrelations for both the squared values of the observed time series
 and the squared residuals estimated from the fitted GARCH(1,2).
 As expected, squared SPY log returns are significantly serially correlated, but the correlogram of squared residuals suggests no significant serial correlations except for
some minor ones at lags 1 and 10. This explains the relatively large $p$-values of the LBQ test.
However, squared residuals can be serially uncorrelated, but dependent, and hence it appears that the tests proposed in this paper are better suited than the Ljung-Box~$Q$ test
in detecting 
the misspecification of the conditional variance specification in this case.

\subsubsection*{Example 2}

In this illustrative example we consider a data set from 
the Caterpillar stock traded on the New York Stock Exchange.
The variable of interest is the daily log return of the Caterpillar stock, defined by $Y_{t}=100(\log P_{t}-\log P_{t-1})$ where $P_t$ is the stock price at time $t$.
The sample contains  $2515$ observations and spans the period Jan 02, 2001 to Dec 31, 2010.
\cite{Tsay:12} analyzed this data set by applying several diagnostic methods, and
 fitted a GARCH(1,1) model
 (see  Table~5.1 in~\citealp{Tsay:12}).
 When we fit a GARCH(1,2) model 
 to this data set, the estimated GARCH(2) coefficient turns out to be statistically insignificant, practically at any level of significance.
 This indicates that, when testing the GARCH(1,2) specification, one component of the true parameter could potentially be a boundary point of the parameter space,
 whereas when the null model is GARCH(1,1) the true parameter could potentially be an interior point.
 Of course we do not have any certainty that this is actually true.
 But, as an illustration, we employ the proposed KS and CvM bootstrap tests to
 test GARCH(1,1) and GARCH(1,2) specifications. 
 For comparison the LBQ test is also considered.
 The $p$-values of the tests are given in Table~\ref{tbl:cat}.
 As expected the tests support both GARCH(1,1) and GARCH(1,2) specifications with large $p$-values.
 In the simulations in the previous section, the LBQ test was undersized when testing for the correct specification.
 Thus, the large $p$-values of the LBQ test in Table~\ref{tbl:cat} are consistent with the simulation~results reported in the previous section.

 \begin{table}[H]
\caption{The $p$-values of the specification tests for testing GARCH(1,1) and GARCH(1,2) specifications
for the conditional variance of the daily log-return of the Caterpillar stock. The data spans the period Jan 02, 2001 to Dec 31, 2010. \vspace{-.8cm}}
\centering
\renewcommand{\arraystretch}{1.1}
\rule[-8pt]{0pt}{22pt}
\begin{tabular*}{\textwidth}{@{\extracolsep{\fill}} l  c c  c c c  c c }
  \hline \hline
  &\multicolumn{7}{c}{Tests}\\ \cline{2-8}\\[-.25cm]
  Null model &  {KS} & {CvM} & {LBQ(3)} & LBQ(5) & LBQ(10) & LBQ(15) & LBQ(20)  \\ \cline{3-4}
  \hline \\ [-.25cm]
  GARCH(1,1) & 0.363 &	0.332 &	0.943 &	0.995 &	0.999 &	0.999 &	0.999 \\[.1cm]
  GARCH(1,2) & 0.491 &	0.427 &	0.880 &	0.978 &	0.999 &	0.998 &	0.999 \\[.1cm]
  \hline \hline \\ [-.35cm]
\end{tabular*}
\label{tbl:cat}
\end{table}


\section{Conclusion}\label{sec:con}

This paper contributes to advance the current statistical 
methodology for inference in GARCH models by developing bootstrap based omnibus specification tests while allowing parameters on the boundary of the parameter space.
In particular, Kolmogorov-Smirnov and Cram\'{e}r-von Mises type test statistics are proposed based on a certain empirical process marked by centered squared residuals.
We first derive the asymptotic null distributions of the proposed test statistics when the true parameter is in the interior of the parameter space. 
Since the limiting distributions of
the test statistics are not free from (unknown) nuisance parameters, we propose a bootstrap method to implement the tests  and
establish that the proposed bootstrap method is asymptotically valid and consistent.
However, when some components of the nuisance parameters lie on the boundary of the parameter space, this bootstrap testing procedure is not consistent.
Hence, as an alternative, we also propose a modified version of the bootstrap 
by employing a method of shrinkage of the parameter estimates in the bootstrap data generation. 
We show that the modified bootstrap procedure is asymptotically valid and consistent, regardless of the presence of nuisance parameters on the boundary.
Our bootstrap methods can be implemented easily under fairly general and easily verifiable assumptions and
have desirable finite sample properties in terms of empirical size and power.


Our results can be extended in several directions. For instance,
it is of interest to see if the methods we propose in this paper can be extended to models beyond the standard GARCH($p_1,p_2$).
To this end, consider the model $\cM$ defined~by
\benr
\label{main:model:g}
\cM: \quad
Y_i &= {\it h}_i^{1/2} \varepsilon_i,\quad
{\it h}_i = g_{\bfphi}(Y_{i-1}, \cdots, Y_{i-p_1}, {\it h}_{i-1}, \cdots, {\it h}_{i-p_2}), \quad i \in \Z,
\eenr
for some $\bfphi \in \Phi \subset \mathbb{R}^{p_1+p_2}$, where $\{g_\bfphi; \bfphi \in \Phi\}$ is a parametric family of nonnegative functions on $\R^{p_1+p_2}$, and
the error terms $\{\varepsilon_i\}_{i \in \mathbb{Z}}$ are i.i.d.\,\,with
zero mean and unit variance. 
Thus,
$
{\it h}_i={\it h}_i(\bfphi)=\var(Y_i\mid \cH_{i-1}), \ i \in \Z.
$
Consider the hypothesis testing problem
\ben
\label{eq-hyp-1g}
\mathsf{H}_0: \mbox{Model } \cM \mbox{ is correct } \quad \mbox{ vs } \quad \mathsf{H}_1: \mbox{Model } \cM \mbox{ is not correct}.
\een
The $\textrm{GARCH($p_1,p_2$)}$ model is a special case of $\cM$.
Another example is the asymmetric AGARCH($p_1,p_2$) model defined by
$
{\it h}_i = {\it h}_i(\bfphi) = \alpha_0 + \sum_{j=1}^{p_1} \alpha_j \big(|Y_{i-j}| - \gamma Y_{i-j} \big)^2 + \sum_{k=1}^q \beta_{k} {\it h}_{i-k},
$
where $\bfphi=(\alpha_0, \ldots, \alpha_p, \beta_1, \ldots, \beta_q, \gamma)^\prime$, $\alpha_0 >0, \alpha_j \geq 0$, $\beta_{k} \geq 0$
 ($i \in \mathbb{Z}$, $1 \leq j \leq p$, $1 \leq k \leq q$).
 Similarly, several other extensions of the standard GARCH model can also be written in the general form~\eqref{main:model:g}.

Heuristic arguments suggest that the bootstrap tests proposed in this paper for ARCH($p$) and  GARCH($p_1,p_2$) models
can also be extended to this general setup.
In fact, the bootstrap algorithm outlined in Section~\ref{sec:boot:new} can be readily applied to any model of the form~\eqref{main:model:g},
based on a suitable estimator for $\bfphi$.
However, parameter estimation,
when the true value is on the boundary of the parameter space, in the family of models in~\eqref{main:model:g}, has not yet been studied, 
and therefore it is not a trivial task to extend the methods developed in this~paper to
a general setup of the form~\eqref{main:model:g}; this would
provide a potential direction for a possible extension of the paper.
Furthermore, our testing procedures can also be potentially extended to Poisson autoregressions with exogenous covariates
as considered in \cite{AGOSTO:16}.


\begin{appendix}

\section{APPENDIX: Assumptions and Proofs}
\label{appendixA}
\renewcommand{\theequation}{A.\arabic{equation}}
\renewcommand{\thesection}{A}
\setcounter{equation}{0}

\setcounter{theorem}{0}\setcounter{lemma}{0}
\renewcommand{\thelemma}{\thesection.\arabic{lemma}}
\renewcommand{\theproposition}{\thesection.\arabic{proposition}}

\subsection{Some notations and assumptions}

In this appendix we introduce some additional notation, and state Assumptions~\ref{A:1}--\ref{A:power} required for the main theorems.
First, we generalize the data generating process specified by model~\eqref{con:var:mod}--\eqref{null1}
for an arbitrary $\bfphi \in \Phi$ and a given innovation distribution~$F$ with zero~mean and unit variance.
To this end we need to first introduce the following regularity assumption.


\begin{Bcondition}\label{A:1}
The process $\{Y_i\}_{ i \in \mathbb{Z}}$ is strictly stationary and ergodic
and obeys model~\eqref{con:var:mod} under the alternative hypothesis $\mathsf{H}_1$.
The parameter space $\Phi$ is a compact subset of $(0, \iny) \times [0,\iny)^{p_1+p_2}$
and contains a hypercube of the form
$[\om_L, \om_U] \times [0, \ep]^{p_1+p_2}$, for some $\ep > 0$ and $\om_U > \om_L > 0$,
which includes $\bfphi_0^* = (\om_0^*, \al_{01}^*, \ldots, \al_{0p_1}^*, \b_{01}^*, \ldots, \b_{0p_2}^*)^\prime$, where $\bfphi_0^*$ is
the pseudo-true parameter value under $\mathsf{H}_1$, defined by $\bfphi_0^*:= \plim \h \bfphi$,
where `$\plim$' is the probability limit as $n \to \iny$.
Further,
$\sum_{i=1}^{p_1} \al_{0i}^* + \sum_{j=1}^{p_2}\b_{0j}^* < 1$ for $p_1 \ge 1$, $p_2 \ge 0$, $\sum_{i=1}^{p_1} \al_{0i}^* \neq 0$.
\end{Bcondition}

The strict stationarity of the process $\{Y_i : i \in \Z\}$ obeying \eqref{con:var:mod}--\eqref{null1},
which follows from \ref{qml:1}, \ref{qml:2} and \ref{qml:4},
 ensures that
the true parameter $\bfphi_0 = (\om_0, \al_{01}, \ldots, \al_{0p_1}, \b_{01}, \ldots, \b_{0p_2})^\prime$
under the null hypothesis $\mathsf{H}_0$ satisfies
$\sum_{i=1}^{p_1} \al_{0i} + \sum_{j=1}^{p_2} \b_{0j} < 1$ (see \citealp{Bougerol:Picard:92,bougerol1992}).
Assumption~\ref{A:1} assumes that this continues to hold
when $\bfphi_0^*$ is the pseudo true value under the alternative hypothesis~$\mathsf{H}_1$.
Since $(\bfphi_0^*,F_0^*):= \plim (\h \bfphi, \check F_n)$, 
under the null hypothesis $\mathsf{H}_0$, we have that $(\bfphi_0^*,F_0^*) = (\bfphi_0,F_0)$, and under the alternative hypothesis $\mathsf{H}_1$,
$(\bfphi_0^*,F_0^*)$ is the pseudo-true value of $(\bfphi,F)$.
Therefore, 
if either 
$\mathsf{H}_0$ holds under~\ref{qml:1}--\ref{qml:4} or
$\mathsf{H}_1$ holds under~\ref{A:1}, 
regardless of whether $\bfphi_0^*$ is in the interior or on the boundary of $\Phi$,
for all sufficiently small $\ep > 0$ and $\om_U > \om_L > 0$,
there exists a hypercube of the form
\ben\label{hcube}
\bar \Phi := [\om_L, \om_U] \times [0, \ep]^{p_1+p_2} \subset \Phi
\een
including $\bfphi_0^*$, such that,
for every $\bfphi \in \bar \Phi$ and c.d.f.\,\,$F$ (with mean 0 and variance 1),
 the model defined~by
\benr\label{DGmodel}
\notag && Y_i^{(\bfphi, F)} = \{{\it h}_i^{(\bfphi, F)}(\bfphi)\}^{1/2}\vep_i^{(F)}, \\
&& {\it h}_i^{(\bfphi, F)}(\bfphi)  = \om+ \sum_{j=1}^{p_1} \al_j \{Y_{i-j}^{(\bfphi, F)}\}^2 + \sum_{j=1}^{p_2} \b_j {\it h}_{i-j}^{(\bfphi, F)}(\bfphi),
\eenr
has a unique strictly stationary and ergodic solution with $\mE[\{Y_0^{(\phi, F)}\}^2] < \iny$, 
where $\vep_i^{(F)} = F^{-1}(U_i):= \inf \{y \in \R : F(y) \ge U_i\}$
and $\{U_i, i \in \mathbb{Z}\}$ are i.i.d.\,uniform(0,1) random variables,
for example, by Theorem 2.1 of \cite{Chen:Hong:98}.

For $(\bfphi, F)=(\bfphi_0, F_0)$ the model~\eqref{DGmodel} is equivalent to the
DGP defined by %
\eqref{con:var:mod}--\eqref{null1}.~Usually,
$\bfphi_0$ and $F_0$ are unknown. Hence, in order to generate data from a model that mimics %
\eqref{con:var:mod}--\eqref{null1},
one needs to replace $(\bfphi_0, F_0)$ by some known $(\bfphi_n, F_n)$ which is sufficiently close to $(\bfphi_0, F_0)$.
Let $(\bfphi_n, F_n)$ be such a sequence %
in the product space $\bar \Phi \times \cD(\R)$, $F_n$ is a c.d.f\,\,with zero mean and unit variance ($n \in \mathbb{N}$),
such that $(\bfphi_n, F_n) \rightarrow (\bfphi_0^*, F_0^*)$ as $n \to \iny$,
with $\|\bfphi_n - \bfphi_0^*\| \ra 0$ and $d_2(F_n, F_0^*)\ra 0$ as $n \to \iny$,
where $d_2(F_X, F_Y)$ denotes the Mallows metric for the distance between two probability distributions
$F_X$ and $F_Y$ (see Lemma~\ref{lem:hf}).
Note that, since $(\bfphi_0^*, F_0^*) = \plim (\h \bfphi, \check F_n)$, we have $(\bfphi_0^*,F_0^*) = (\bfphi_0,F_0)$ under $\mathsf{H}_0$,   and
$(\bfphi_0^*,F_0^*)$ is the pseudo-true value   under $\mathsf{H}_1$.
In what follows, when the DGP~\eqref{DGmodel} corresponds to $(\bfphi_n, F_n)$
instead of using ${\it h}_i^{(\phi_n, F_n)}(\cdot)$
and $\tau_i^{(\phi_n, F_n)}(\cdot)$, we let the analogs of~${\it h}_i(\cdot)$
and $\tau_i(\cdot)$ be denoted by ${\it h}_{ni}(\cdot)$ and $\tau_{ni}(\cdot)$, respectively.
Note that,
under $\mathsf{H}_0$, the probability laws of ${\it h}_{i}^{(\phi_0, F_0)}(\cdot)$ and $\tau_{i}^{(\phi_0, F_0)}(\cdot)$
are identical~to those of ${\it h}_{i}(\cdot)$ and $\tau_{i}(\cdot)$, respectively.

Next, let us introduce some notation. Let `dot' denote differentiation:
\benn
\text{$\dot{{\it h}}_i(\bfphi) = (\partial/\partial \bfphi) {\it h}_i(\bfphi)$, \quad
$\ddot{{\it h}}_i(\bfphi) = (\partial/\partial \bfphi) \dot{{\it h}}_i(\bfphi)$.
 }
\eenn
Let $\cF$ denote the set of all c.d.f.'s with zero mean and unit variance, i.e,
\benn
\cF := \{ F \in \cD(\R) : \text{$F$ is a c.d.f. with mean 0 and variance 1}\}.
\eenn
For any given constant  $\del > 0$, let
$
\cF_\del^0 := \{ F \in \cF : d_2(F, F_0^*) \le \del\}.
$

We say that a sequence of random variables $\{Z_i\}_{i \in \mathbb{N}}$ \textit{converges to zero exponentially almost surely},
denoted $Z_i \stackrel{e.a.s.}{\rightarrow} 0$, if
there exists $\gamma >1$ such that $\gamma^i Z_i \stackrel{a.s.}{\rightarrow} 0$ as~$i \rightarrow \infty.$
The norm $\|\cdot\|_\Lambda$ for
a continuous matrix-valued function~$H$ on a compact set $\Lambda \subset \R^{r_1}$, that is
$H \in \mathbb{C} [\Lambda, \R^{r_2\times r_3}]$, is defined~by
$
\|H\|_\Lambda := \sup_{s \in \Lambda}\|H(s)\|,
$
when $r_1, r_2, r_3$ are known positive integers.
If $H$ is real valued, 
then $\|H\|_\Lambda = \sup_{s \in \Lambda}|H(s)|.$
We let $\bar \R := [-\iny,\iny]$.

In order to establish the asymptotic validity of the bootstrap testing procedure 
we also introduce the following additional assumptions. Recall that $(\bfphi_0^*, F_0^*) = \plim (\h \bfphi, \check F_n)$.

\begin{Bcondition}\label{A:1:c}
There exist $\del > 0$ with
$\mE([\sup_{F \in \cF_\del^0} \{F^{-1}(U_i)\}^2]^{2+d})< \infty$
for some   $d >0$.
\end{Bcondition}

\begin{Bcondition}\label{A:1:c:n}
If $\|\bfphi_n - \bfphi_0^*\| \ra 0$ and $d_2(F_n, F_0^*)\ra 0$ as $n \to \iny$, then~for every $y \in \bar \R$, we have that
$\mE \mathbb{I}(Y_1^{(\bfphi_n, F_n)} \leq y) \ra \mE \mathbb{I}(Y_1^{(\bfphi_0^*, F_0^*)} \leq y)$ as $n \to \iny$.
\end{Bcondition}

\begin{Bcondition}\label{A:3:c}
For every nonrandom sequence $\zeta_n :=(\bfphi_n, F_n) \ra \zeta_0^*:= (\bfphi_0^*, F_0^*)$, where $\zeta_n \in \bar \Phi \times \cF$, we have that
$\mE[\tau_{n1}(\bfphi_n)] \ra \mE[ \tau_1^{(\phi_0^*, F_0^*)}(\bfphi_0^*)]$,
$\mE[\tau_{n1}(\bfphi_n)\tau_{n1}(\bfphi_n)^\prime] \ra \mE [\tau_1^{(\phi_0^*, F_0^*)}(\bfphi_0^*)\tau_1^{(\phi_0^*, F_0^*)}(\bfphi_0^*)^\prime]$ as $n \to \iny$.
Recall that
$
\tau_{ni}(\bfphi) := \dot{{\it h}}_{ni}(\bfphi)/{\it h}_{ni}(\bfphi)$ with ${\it h}_{ni}(\bfphi) = {\it h}_i^{(\bfphi_n, F_n)}(\bfphi).
$
\end{Bcondition}

The next assumption is used in the proof of Theorem~\ref{thm:power} to establish the consistency of the bootstrap test (\ref{eq:boot:tst}) based on $T_j$ ($j = 1, 2$).

\begin{Bcondition}\label{A:power}
There exists a $y \in \R$, with ${\it h}_i = \mE(Y_i^2 \mid \cH_{i-1})$, $i \in \Z$, such that
$\mE[\{{\it h}_1/{\it h}_1(\bfphi_0^*) - 1\}\mathbb{I}(Y_0 \le y)] \neq 0$ under $\mathsf{H}_1$, where  $\bfphi_0^*$ is the pseudo-true value 
under $\mathsf{H}_1$.
\end{Bcondition}

\subsection{Some preliminary results}

In this subsection we obtain several preliminary lemmas required for the main proofs.

The next lemma shows that $\check F_n$ in~\eqref{checkF} converges to $F_0$ with probability~1.

\begin{lemma}\label{lem:hf}
Let $d_2(F_X, F_Y)$ denote the Mallows metric for the distance between two probability distributions
$F_X$ and $F_Y$ defined by $d_2(F_X, F_Y) = \inf\{\mE|X-Y|^2\}^{1/2}$, where the infimum is over all square integrable random variables $X$ and $Y$ with marginal distributions $F_X$ and $F_Y$.
\emph{(a)} Suppose that Assumptions~\ref{qml:1}--\ref{qml:5} and $\mathsf{H}_0$ hold, and $\bfphi_0$ is an interior point in~$\Phi$.
Then, $d_2(\check F_n, F_0)\stackrel{a.s.}{\rightarrow} 0$ as $n \to \iny$.
\emph{(b)} Additionally, assume that Assumption~\ref{qml:6} is also satisfied, then
$d_2(\check F_n, F_0)\stackrel{a.s.}{\rightarrow} 0$ as $n \to \iny$, irrespective of whether $\bfphi_0$ is in the interior of $\Phi$.
\end{lemma}

\begin{proof}[\bf Proof of Lemma~\ref{lem:hf}]
Under Assumptions~\ref{qml:1}--\ref{qml:4}, $\h \bfphi$ converges to $\bfphi_0$ (a.s.), 
irrespective of whether $\bfphi_0$ is in the interior of the parameter space (see Lemma~\ref{thm:1g}).
If, in addition, \ref{qml:5} is also satisfied and $\bfphi_0$ is in the interior of $\Phi$, then
$\h \bfphi$ is asymptotically linear and satisfies~\eqref{th:expan}, and hence $n^{1/2}(\h \bfphi - \bfphi_0) = O_p(1)$.
If Assumptions~\ref{qml:1}--\ref{qml:6} are satisfied, then $n^{1/2}(\h \bfphi - \bfphi_0) = O_p(1)$ by Lemma~\ref{thm:1g}, irrespective of whether $\bfphi_0$ is in the interior of $\Phi$.

\smallskip

\noi \emph{Proof of Part (a): Assumptions~\ref{qml:1}--\ref{qml:5} are satisfied, and $\bfphi_0$ is an interior point in $\Phi$.}

\smallskip

Let $H_n(x) := n^{-1} \s \mathbb{I}(\varepsilon_i \leq x)$, be the empirical distribution function of the unobserved errors $\{{\varepsilon}_1,\ldots,{\varepsilon}_n\}$.
From the triangular inequality we have that
\benn
d_2(\wh F_n, F_0) \le d_2(\wh F_n, H_n) + d_2(H_n, F_0),
\eenn
where $H_n(x) := n^{-1} \s \mathbb{I}(\varepsilon_i \leq x)$, $x \in \R$, is the empirical distribution function of the unobserved errors $\{{\varepsilon}_1,\ldots,{\varepsilon}_n\}$.

We already have that $d_2(H_n, F_0) \stackrel{a.s.}{\rightarrow} 0$ as $n \to \iny$
(see, for example, Lemma~8.4 of \citealp{Bickel:Freedman:81}).
Thus, it suffices to show that $d_2(\check F_n, H_n) \stackrel{a.s.}{\rightarrow} 0$ as $n \to \iny$.
To this end, let $J$ be a random variable having Laplace distribution on $\{1, \ldots, n\}$,
with $P(J=i) = 1/n$ for each $i = 1, \ldots, n.$
Define two random variables $X^{(1)}$ and $Y^{(1)}$ by
\benn
\text{$X^{(1)} = \vep_J$ and $Y^{(1)} = \h \vep_J$.}
\eenn
Then, $X^{(1)}$ and $Y^{(1)}$ have the marginal distributions $H_n$ and $\check F_n$ respectively.
Therefore, 
\benr
\notag \lefteqn{\{d_2(\check F_n, H_n)\}^2 = \inf\{\mE|X-Y|^2\} \le \mE\{X^{(1)}-Y^{(1)}\}^2}\\ \label{eq:hf:1}
&&= n^{-1} \s (\vep_i - \check \vep_i)^2
= (n\h \sigma_n^2)^{-1} \s \{\h \sigma_n \vep_i - (\wh \vep_i - n^{-1} \sum_{j=1}^n \wh \vep_j)\}^2,
\eenr
where $\h \sigma_n^2 = n^{-1} \s \{\wh \vep_i - n^{-1} \sum_{j=1}^n \wh \vep_j\}^2$.

Since $\h \bfphi \stackrel{a.s.}{\rightarrow} \bfphi_0$ and 
$n^{1/2}(\h \bfphi - \bfphi_0) = O_p(1)$,
it follows %
that $\h \sigma_n^2 \stackrel{a.s.}{\rightarrow} 1$.
Hence, for some constant $K>0$, \eqref{eq:hf:1} is bounded from above by
\ben\label{eq:hf:2}
Kn^{-1}\s (\wh \vep_i - \vep_i)^2 + Kn^{-2}\bigg( \s \vep_i \bigg)^2 + M_n
\een
where $M_n$ is a random variable that converges to zero with probability one.
Here we have used some arguments from the proof of Lemma~6 in \cite{Perera:Silvapulle:20}.

Furthermore, for some $K < \iny$, we have that
\benrr
n^{-1}\s (\wh \vep_i - \vep_i)^2
&\le& Kn^{-1}\s \vep_i^2\big[\{{\it h}_i(\bfphi_0)\}^{1/2} - \{{\it h}_i(\h \bfphi)\}^{1/2} \big]^2.
\eenrr
From Proposition~\ref{prop:A4:A9}, we obtain that
\ben\label{rooth:con}
\{{\it h}_i(\h \bfphi)\}^{1/2} - \{{\it h}_i(\bfphi_0)\}^{1/2} = 2^{-1}(\h \bfphi - \bfphi_0)^\prime \dot{{\it h}}_i(\bfphi_0)/ \{{\it h}_i(\bfphi_0)\}^{1/2} + o_p(n^{-1/2}).
\een
Because $\h \bfphi \stackrel{a.s.}{\rightarrow} \bfphi_0$ and $\|\mE\vep_1^2\dot{{\it h}}_i(\bfphi_0)/ \{{\it h}_i(\bfphi_0)\}^{1/2}\| < \iny$, then we have that
\benn
\text{$n^{-1} \s \vep_i^2[\{{\it h}_i(\h \bfphi)\}^{1/2} - \{{\it h}_i(\bfphi_0)\}^{1/2}] \stackrel{a.s.}{\rightarrow} 0$ as $n \to \iny$.}
\eenn
Consequently, $d_2(\check F_n, H_n) \stackrel{a.s.}{\rightarrow} 0$ and hence $d_2(\check F_n, F_0)\stackrel{a.s.}{\rightarrow} 0$.

\smallskip

\noi \emph{Proof of part (b):} 
Under Assumptions~\ref{qml:1}--\ref{qml:6}, from Proposition~\ref{prop:A4:A9} below, we obtain that \eqref{rooth:con} holds irrespective of whether $\bfphi_0$ is in the interior of $\Phi$.
Since $\h \bfphi \stackrel{a.s.}{\rightarrow} \bfphi_0$, and $n^{1/2}(\h \bfphi - \bfphi_0) = O_p(1)$, then it follows that $d_2(\check F_n, F_0) \stackrel{a.s.}{\rightarrow} 0$ by repeating the arguments
of the proof of part~(a).
\end{proof}

\begin{lemma}\label{lem:A:1}
Suppose that either Assumption~\ref{qml:1} or Assumption~\ref{A:1} is satisfied.
Additionally, assume that Assumption~\ref{A:1:c} holds.
Then, for every $\zeta=(\bfphi, F) \in \bar \Phi \times \cF_\del^0$
\begin{description}
  \item[a)] The model~\eqref{DGmodel} has a unique stationary ergodic solution
$\{Y_i^{(\zeta)} : i \in \mathbb{Z} \}$,
  \item[b)] $\mE [|Y_i^{(\zeta)}|^{4+d}]$, $\mE[|{\it h}_{i}^{(\zeta)}(\bfphi)|^{2+d}]$ and $\mE[\|\tau_i^{(\zeta)}(\bfphi)\|^{2+d}]$ are finite for some $d > 0$.
\end{description}
\end{lemma}

\begin{proof}
Let $\zeta=(\bfphi, F) \in \bar \Phi \times \cF_\del^0$ be fixed and arbitrary.
Since $F$ has zero mean and unit variance, the condition
$\sum_{i=1}^{p_1} \al_i + \sum_{j=1}^{p_2} \b_j < 1$
is necessary and sufficient for the process $\{Y_i^{(\zeta)}; i \in \Z\}$
to be strictly stationary and have finite second moments with $\mE(Y_i^{(\zeta)}) = 0$ and $\mE[\{Y_i^{(\zeta)}\}^2] = \om / ( 1 - \sum_{i=1}^{p_1} \al_i - \sum_{j=1}^{p_2} \b_j$);
see for example \cite{Nelson:90} and \cite{Chen:Hong:98}. 
Therefore, if $\mathsf{H}_0$ holds, then part~(a) follows from Assumption~\ref{qml:1}, and if $\mathsf{H}_1$ holds, then it follows from Assumption~\ref{A:1}.

Since $\vep_i^{(F)} = F^{-1}(U_i)$ 
with $\{U_i, i \in \mathbb{Z}\}$ being i.i.d.\,\,uniform(0,1) random variables, from Assumption~\ref{A:1:c}, we have
$\mE([\{\vep_i^{(F)}\}^2]^{2+d})< \infty$, and hence part~(b) also follows.
\end{proof}

\begin{lemma}\label{lem:A:3:c}
Suppose that either Assumption~\ref{qml:1} or Assumption~\ref{A:1} is satisfied.
Additionally, assume that Assumptions~\ref{A:1:c}--\ref{A:3:c} hold.
Then, for every nonrandom sequence $\zeta_n :=(\bfphi_n, F_n) \ra \zeta_0^*:= (\bfphi_0^*, F_0^*)$, where $\zeta_n \in \bar \Phi \times \cF$, we have that
\ben\label{th:expan2}
\mid n^{1/2}(\h \bfphi_{nn} -\bfphi_n) - \Si_{nn}^{-1}(\bfphi_n) n^{-1/2}\s \tau_{ni}(\bfphi_n)(\vep_{ni}^2-1) \mid = o_p(1),
\een
with
$
\Si_{nn}(\bfphi) := n^{-1} \s \tau_{ni}(\bfphi) \tau_{ni}(\bfphi)^\prime
$,
and $\vep_{ni} = F_n^{-1}(U_i)$,
where
$\h \bfphi_{nn}$~is~the analogue of $\h \bfphi$ for the data generating process at~$(\bfphi_n, F_n)$, and
$\bfU = \{U_i, i \in \mathbb{Z}\}$ denote a sequence of i.i.d. random variables from the uniform(0,1) distribution.
\end{lemma}

\begin{proof}
The proof follows from  arguing as in the proof of Lemma~4(a) in \cite{Perera:Silvapulle:21} for the GARCH($p_1,p_2$) setup.
\end{proof}

\begin{lemma}\label{lem:A4:A9}
Suppose that either Assumption~\ref{qml:1} or Assumption~\ref{A:1} is satisfied.
Additionally, assume that Assumptions \ref{A:1:c}--\ref{A:3:c} hold.
Then, for every constant $C < \iny$, 
\newline {\em i)} \, $\sup | {\it h}_{ni}^{1/2} (\bfb) - {\it h}_{ni}^{1/2}(\bfa)- 2^{-1}(\bfb-\bfa)^{\prime}\dot{{\it h}}_{ni}(\bfa){\it h}_{ni}^{-1/2}(\bfa) |{\it h}_{ni}^{-1/2}(\bfa) = o_{p}(n^{-1/2}),$
\newline {\em ii)} \, $\sup  \mid {\it h}_{ni}(\bfb) - {\it h}_{ni}(\bfa) - (\bfb-\bfa)^{\prime}\dot{{\it h}}_{ni}(\bfa)\mid {\it h}_{ni}^{-1}(\bfa) = o_{p}(n^{-1/2})$,\\
where the supremum is taken over $1 \le i \le n$ and over
$\{(\bfb,\bfa) : \bfb, \bfs \in \bar \Phi, \sqrt{n}\|\bfb - \bfa\| \le C\}$.
\end{lemma}

\begin{proof}[Proof of Lemma~\ref{lem:A4:A9}]
Let $D := \bar \Phi \times \cF_\del^0$, $p = p_1+p_2+1$.
Let $\wp_{ni}(\bfphi) = \{{\it h}_{ni}(\bfphi)\}^{1/2}$ for $\bfphi \in \bar \Phi$.
Let $\Delta_{ni}(\bfa, \bfb) := {\it h}_{ni}^{1/2} (\bfb) - {\it h}_{ni}^{1/2}(\bfa)- 2^{-1}(\bfb-\bfa)^{\prime}\dot{{\it h}}_{ni}(\bfa){\it h}_{ni}^{-1/2}(\bfa)$.
Let $\bfa, \bfb \in \bar \Phi$ be fixed but arbitrary. Then, for each $n \in \mathbb{N}$, there exists $\bfdelta_{n1} \in \R^p$,
such that
$\bfb = \bfa + n^{-1/2}\bfdelta_{n1}$.
Hence,
$\Delta_{ni}(\bfa, \bfb) = \wp_{ni}(\bfb) - \wp_{ni}(\bfa)- n^{-1/2}\bfdelta_{n1}^{\prime}\dot{\wp}_{ni}(\bfa).$
Therefore,
by the Mean Value Theorems for functions from $\R^p$ to $\R$, and $\R^p$ to $\R^p$, with right partial derivatives,
for every $n \in \mathbb{N}$,
there exist $\bfdelta_{n2}, \bfdelta_{n3} \in \R^p$ with $\|\bfdelta_{n3}\| \le \|\bfdelta_{n2}\| \le \|\bfdelta_{n1}\|$, such that
\benrr
\Delta_{ni}(\bfa, \bfb) \ = \ n^{-1/2}\bfdelta_{n1}^{\prime}\left[\dot{\wp}_{ni}(\bfa+n^{-1/2}\bfdelta_{n2}) - \dot{\wp}_{ni}(\bfa)\right]
\ = \ n^{-1}\bfdelta_{n1}^{\prime}\ddot{\wp}_{ni}(\bfa+n^{-1/2}\bfdelta_{n3})\bfdelta_{n2},
\eenrr
where
\benn
\ddot{\wp}_{ni}(\bfphi) = \frac{1}{2}\frac{\ddot{{\it h}}_{ni}(\bfphi)}{\{{\it h}_{ni}(\bfphi)\}^{1/2}} -
\frac{1}{4}\frac{\tau_{ni}(\bfphi)\ddot{{\it h}}_{ni}^\prime(\bfphi)}{\{{\it h}_{ni}(\bfphi)\}^{1/2}}. 
\eenn
Therefore, for any given constant $C > 0$, w.p.\,1,
\ben\label{d:bound}
\max_{1 \le i \leq n}\sup_{\bfa, \bfb \in \bar \Phi,
   {\sqrt{n}\|\bfb - \bfa\| \leq C}} n^{1/2}|\Delta_{ni}(\bfa, \bfb)| \ \le \ n^{-1/2}C^2\max_{1 \le i \leq n}\|\ddot{\wp}_{ni}\|_{\bar \Phi}. 
\een
For example, for the GARCH(1,1) case, by differentiation, we have
$$\dot{{\it h}}_{ni}(\bfphi) = \{(1 - \b)^{-1}, \sum_{j=1}^{\infty} \b^{j-1} Y_{n (i-j)}^2,
\om (1 - \b)^{-2} + \al \sum_{j=2}^{\infty} (j-1) \b^{j-2} Y_{n (i-j)}^2\}^\prime,$$
and the $3 \times 3 $ matrix $\ddot{{\it h}}_{ni}(\bfphi)$ is given by
\benrr
&&\text{$[\ddot{{\it h}}_{ni}]_{rk}(\bfphi) = 0 $ for $r=1,2,$ $k =1,2$, $\quad [\ddot{{\it h}}_{ni}]_{13}(\bfphi) =  [\ddot{{\it h}}_{ni}]_{31}(\bfphi) = (1 - \b)^{-2}$,}\\
&&\text{$[\ddot{{\it h}}_{ni}]_{23}(\bfphi) =  [\ddot{{\it h}}_{ni}]_{32}(\bfphi) = \sum_{j=2}^{\infty} (j-1) \b^{j-2} Y_{n (i-j)}^2, \quad$ and}\\
&&\text{$[\ddot{{\it h}}_{ni}]_{33}(\bfphi) = 2\om (1 - \g)^{-3} + \al \sum_{j=3}^{\infty} (j-2) (j-1) \b^{j-3} Y_{n (i-j)}^2.$}
\eenrr
One can similarly obtain the derivatives for the GARCH($p_1,p_2$) for any given $p_1,p_2$.

Since $\sup_{\zeta \in D}\|{\it h}_{i}^{(\zeta)}(\cdot)\|_{\bar \Phi} > \om_L > 0$,
then part~(i) follows under~\ref{A:1:c}--\ref{A:3:c}, 
by arguing as in the proof of Lemma~4 in \cite{Perera:Silvapulle:20} applying Chebyshev's inequality.
The part~(ii) follows~similarly under~\ref{A:1:c}--\ref{A:3:c}.
\end{proof}

The next lemma follows from the proof of Theorem 13.1 in \cite{Billingsley:68} and an application of the
Cauchy-Schwarz inequality; see also Lemma 5.1 in \cite{Stute:97}.
We restate this result here for the ease of reference.
This lemma is useful for establishing the tightness of certain processes.

\begin{lemma}\label{lem:billingsley}
Let $\{(a_i, b_i); 1 \le i \le n\}$ be i.i.d.\,\,square-integrable bivariate
random vectors with $\mE(a_i)=\mE(b_i) = 0$, $1 \le i \le n$. Then we have that
$
\mE\{(\s a_i)^2(\sum_{j=1}^n b_j)^2\} \le n\mE(a_1^2b_1^2)+3n(n-1)\mE(a_1^2)\mE(b_1^2).
$
\end{lemma}

For the proof of Lemma~\ref{thm:1} we make use of the following proposition.
\begin{proposition}\label{prop:A4:A9}
Suppose that the assumptions of Lemma~\ref{thm:1} hold.
Then, for every 
$K < \iny$, 
\newline (a)~
$n^{1/2} \sup \mid {\it h}_{i}^{1/2} (t) - {\it h}_{i}^{1/2}(s)- 2^{-1}(t-s)^{\prime}\dot{{\it h}}_{i}(s){\it h}_{i}^{-1/2}(s) \mid {\it h}_{i}^{-1/2}(\bfphi_0) = o_p(1),$
\newline (b)~ $n^{1/2} \sup  \mid {\it h}_{i}(t) - {\it h}_{i}(s) - (t-s)^{\prime}\dot{{\it h}}_{i}(s)\mid {\it h}_{i}^{-1}(\bfphi_0) = o_p(1)$,
where the supremum is taken over $1 \le i \le n$ and over
$\{(t,s) : t, s \in \varPhi, \sqrt{n}\|t - s\| \le K\}$.
\end{proposition}

\begin{proof}
The proof follows as a special cases of the
proof of Lemma~\ref{lem:A4:A9}.
\end{proof}

\subsection{Initialization effect} 

In this section we obtain several technical results to establish that the effect of initialization in the bootstrap data generation is asymptotically negligible.
First let us introduce some~notation.

\smallskip

\noi \textbf{Notation A:}
If data are generated from~\eqref{DGmodel} for $i\ge -m$,
conditional on a vector of starting values $\varsigma_0 = (y_0, \cdots, y_{1-q}, s_0, \cdots, s_{1-p})^\prime$,
then we %
use the superscript ``$(m,\zeta)$" instead of ``$(\zeta)$", where $\zeta = (\bfphi, F)$.
For example, ${\it h}_{i}^{(m,\zeta)}(\cdot)$ and $\tau_{i}^{(m,\zeta)}(\cdot)$
are the analogues of ${\it h}_i^{(\zeta)}(\cdot)$ and $\tau_{i}^{(\zeta)}(\cdot)$, respectively,
when the data generating model obeys~\eqref{DGmodel} for $i \ge -m$, conditional on the starting values~$\varsigma_0$. 

\begin{lemma}\label{A:new:g}
Suppose that either Assumption~\ref{qml:1} or Assumption~\ref{A:1} is satisfied.
Additionally, assume that Assumption~\ref{A:1:c} holds.
Then, there exists a compact
set $K_1 \subseteq \bar \Phi$, which contains $\bfphi_0^*$, such that the following hold for some $\del > 0$ with $K = K_1 \times \cF_\del^0$:
\begin{description}
  \item[a)] 
  $
\sup_{\zeta \in K} \|{\it h}_{i}^{(m,\zeta)} - {\it h}_{i}^{(\zeta)}\|_{K_1}, \,
\sup_{\zeta \in K} \|\dot {{\it h}}_{i}^{(m,\zeta)} - \dot {{\it h}}_{i}^{(\zeta)}\|_{K_1}, \,
 \stackrel{e.a.s.}{\rightarrow} 0$ as $i \rightarrow \infty$;
  \item[b)] $\sup_{\zeta \in K} \mE\|\ddot{{\it h}}_{0}^{(\zeta)}\|_{K_1}^{2+d}$ and
$\sup_{\zeta \in K} \mE\|\tau_{0}^{(\zeta)}\|_{K_1}^{2+d}$ are finite for some $d>0$.
\end{description}
\end{lemma}

\begin{proof}
The proof follows from  arguing as in the verifications of Conditions (M1) and~(M2)  in \cite{Perera:Silvapulle:21} for the GARCH($p_1,p_2$) setup,
with the set $K_\bftheta$ being replaced by~$\cF_\del^0$.
Sine everything follows after suitable modifications of the arguments already developed in  \cite{Perera:Silvapulle:21}, we omit the details.
\end{proof}

In view of Notation A above, conditional on $(Y_1, \ldots, Y_n)$, we have
\ben\label{boot:op}
\{Y_i^*, {\it h}_i^*(\bfphi), \tau_i^*(\bfphi)\} \equiv \{Y_i^{(0,\zeta^*)}, {\it h}_i^{(0,\zeta^*)}(\bfphi), \tau_i^{(0,\zeta^*)}(\bfphi)\}, \quad \bfphi \in \Phi, i \in \mathbb{N},
\een
where $\zeta^* = (\h \bfphi, \check F_n)$ for the bootstrap method in Section~\ref{sec:boot:test}, and $\zeta^* = (\h \bfphi^\dag, \check F_n)$ for the bootstrap method in Section~\ref{sec:boot:new}
with $\h \bfphi^\dag$ given by~\eqref{shrink:th}.
Similarly, let the bootstrap process
generated by~\eqref{DGmodel}, conditional on $(Y_1, \ldots, Y_n)$,  without any initialization, be defined as
\ben\label{boot:hyp}
\{Y_i^{*(\iny)}, {\it h}_i^{*(\iny)}(\bfphi), \tau_i^{*(\iny)}(\bfphi)\} = \{Y_i^{(\zeta^*)}, {\it h}_i^{(\zeta^*)}(\bfphi), \tau_i^{(\zeta^*)}(\bfphi)\}, \ \bfphi \in \Phi, i \in \Z,
\een
where $\zeta^* = (\h \bfphi, \check F_n)$ and $\zeta^* = (\h \bfphi^\dag, \check F_n)$ for the bootstraps in Section~\ref{sec:boot:test} and Section~\ref{sec:boot:new}, respectively.
Thus, $\{Y_i^{*(\iny)}, {\it h}_i^{*(\iny)}(\bfphi), \tau_i^{*(\iny)}(\bfphi)\}$ %
represents a hypothetical (non-operational) bootstrap process generated by~\eqref{DGmodel}, without any initialization.
Further, let
\benrr
&&\h \bfphi^{*(\iny)} = \arg\min_{\bfphi \in \Phi} \s \ell_i^{*(\iny)}(\bfphi),
 \quad \ell_i^{*(\iny)}(\bfphi) = \log {\it h}_i^{*(\iny)}(\bfphi) + \frac{\{Y_i^{*(\iny)}\}^{2}}{{\it h}_i^{*(\iny)}(\bfphi)},\\
&&\mathcal{U}_n^{*(\iny)}(y,\bfphi)
= n^{-1/2} \sum_{i=1}^n \bigg(\frac{\{Y_i^{*(\iny)}\}^2}{{\it h}_i^{*(\iny)}(\bfphi)} - 1\bigg) \mathbb{I}(Y_{i-1}^{*(\iny)} \leq y),
\quad y \in \R, \, \bfphi \in \Phi.
\eenrr
Since no initialization is used in the bootstrap data generation in~\eqref{boot:hyp}, the marked empirical process
$\mathcal{U}_n^{*(\iny)}(y,\h \bfphi^{*(\iny)})$, unlike $\mathcal{U}_n^{*}(y,\h \bfphi^*)$, is not subject to any initialization error.

The next lemma shows that, conditional on $(Y_1, \ldots, Y_n)$, for the bootstrap method in Section~\ref{sec:boot:test},
$\mathcal{U}_n^{*}(y,\h \bfphi^*)$ and $\mathcal{U}_n^{*(\iny)}(y,\h \bfphi^{*(\iny)})$ are uniformly close, in probability.

\begin{lemma}\label{lem:hyp:boot}
Suppose that the assumptions of Theorem~\ref{thm:3} are satisfied with $\bfphi_0$ being an interior point in $\Phi$, or the assumptions of Theorem~\ref{thm:power} are satisfied.
Then, conditional on $(Y_1, \ldots, Y_n)$, 
 $\sup_{y\in \R}\mid \mathcal{U}_n^{*}(y,\h \bfphi^*) - \mathcal{U}_n^{*(\iny)}(y,\h \bfphi^{*(\iny)}) \mid = o_{p}^{\ast }(1),$ in probability.
\end{lemma}

\begin{proof}
Assume without loss of generality that $Y_i^* \le Y_i^{*(\iny)}$. Then, we have
\benr\label{i+ii}
\nn \lefteqn{\mathcal{U}_n^{*}(y,\h \bfphi^*) - \mathcal{U}_n^{*(\iny)}(y,\h \bfphi^{*(\iny)})}\\
\nn &=& \ n^{-1/2} \sum_{i=1}^n \bigg(\frac{\{Y_i^{*}\}^2}{{\it h}_i^{*}(\h \bfphi^{*})}  - \frac{\{Y_i^{*(\iny)}\}^2}{{\it h}_i^{*(\iny)}(\h \bfphi^{*(\iny)})}\bigg) \mathbb{I}(Y_{i-1}^{*(\iny)} \leq y)\\
\nn && + \  n^{-1/2} \sum_{i=1}^n \bigg(\frac{\{Y_i^{*}\}^2}{{\it h}_i^{*}(\h \bfphi^{*})} - 1\bigg) \mathbb{I}(Y_{i-1}^* \le y < Y_{i-1}^{*(\iny)})\\
&=& \textbf{I} + \textbf{I}\textbf{I}, \quad \text{say.}
\eenr
Since Lemma~\ref{A:new:g}(a) shows that $|{\it h}_{i}^*(\h \bfphi) - {\it h}_{i}^{*(\iny)}(\h \bfphi) | \stackrel{e.a.s.}{\rightarrow} 0$ as $i \rightarrow \infty$,
from Lemma~2.1 in \cite{Straumann:Mikosch:06}, 
$
|\{{\it h}_{i-1}^*(\h \bfphi)\}^{1/2} -\{{\it h}_{i-1}^{*(\iny)}(\h \bfphi)\}^{1/2}| \stackrel{e.a.s.}{\rightarrow} 0, \ \text{as $i \rightarrow \infty$,}
$
and hence from Lemma~2.3 in \cite{Straumann:Mikosch:06}, it follows that
\benn
|Y_{i-1}^* - Y_{i-1}^{*(\iny)}| = |\vep_{i-1}^*| |\{{\it h}_{i-1}^*(\h \bfphi)\}^{1/2} -\{{\it h}_{i-1}^{*(\iny)}(\h \bfphi)\}^{1/2}| \stackrel{e.a.s.}{\rightarrow} 0, \quad  \text{as $i \rightarrow \infty$},
\eenn
and hence the sum \textbf{I}\textbf{I} in \eqref{i+ii} is of order $o_{p}^{\ast }(1),$ in probability, uniformly in $y \in \R$.

The first sum in \eqref{i+ii} is bounded as %
\benrr
|\textbf{I}| &\le&  n^{-1/2} \sum_{i=1}^n \bigg|\frac{\{Y_i^{*}\}^2}{{\it h}_i^{*}(\h \bfphi^{*})}  - \frac{\{Y_i^{*(\iny)}\}^2}{{\it h}_i^{*(\iny)}(\h \bfphi^{*(\iny)})}\bigg| \\
&\le& n^{-1/2} \sum_{i=1}^n \{Y_i^{*}\}^2\bigg|\frac{1}{{\it h}_i^{*}(\h \bfphi^{*})} - \frac{1}{{\it h}_i^{*(\iny)}(\h \bfphi^{*(\iny)})}\bigg| 
 + n^{-1/2} \om_L^{-1}\sum_{i=1}^n \big|\{Y_i^{*}\}^2 - \{Y_i^{*(\iny)}\}^2\big|\\
&=& \textbf{I}_A + \textbf{I}_B, \quad \text{say.}
\eenrr
Since $|\{Y_i^{*}\}^2 - \{Y_i^{*(\iny)}\}^2| = |\vep_{i}^*|^2 |{\it h}_{i}^*(\h \bfphi) - {\it h}_{i}^{*(\iny)}(\h \bfphi)| \stackrel{e.a.s.}{\rightarrow} 0,$ as $i \rightarrow \infty$,
$\sum_{i=1}^n |\{Y_i^{*}\}^2 - \{Y_i^{*(\iny)}\}^2|$ converges a.s.\,\,as $n \to \iny$, by Lemma~2.1 in \cite{Straumann:Mikosch:06},
and hence sum $\textbf{I}_B$ is of order $o_{p}^{\ast }(1)$.

From the proof of Lemma~8 in \cite{Perera:Silvapulle:15a}, $n^{1/2}(\h \bfphi^{*} - \h \bfphi^{*(\iny)}) = o_{p}^{\ast }(1)$.
By Lemma~\ref{A:new:g}, $\sup_{\phi \in K_1}|{\it h}_{i}^*(\bfphi) - {\it h}_{i}^{*(\iny)}(\bfphi) | \stackrel{e.a.s.}{\rightarrow} 0$ for some compact 
set $K_1 \subseteq \bar \Phi$.
Hence, Lemma~2.3 in \cite{Straumann:Mikosch:06} yields that sum $\textbf{I}_A$ is also of order $o_{p}^{\ast }(1)$.
Therefore, $\sup_{y\in \R}\mid \mathcal{U}_n^{*}(y,\h \bfphi^*) - \mathcal{U}_n^{*(\iny)}(y,\h \bfphi^{*(\iny)}) \mid = o_{p}^{\ast }(1),$ in probability.
\end{proof}

The next lemma shows that, conditional on $(Y_1, \ldots, Y_n)$, for the bootstrap method in Section~\ref{sec:boot:new},
$\mathcal{U}_n^{*}(y,\h \bfphi^*)$ and $\mathcal{U}_n^{*(\iny)}(y,\h \bfphi^{*(\iny)})$ are uniformly close, in probability.

\begin{lemma}\label{lem:hyp:boot:2}
Suppose that either assumptions of Theorem~\ref{thm:6} or  Theorem~\ref{thm:power:2} are satisfied.~Then,
 conditional on $\{Y_1, \ldots, Y_n\}$, 
 $\sup_{y\in \R} \mid \mathcal{U}_n^{*}(y,\h \bfphi^*) - \mathcal{U}_n^{*(\iny)}(y,\h \bfphi^{*(\iny)}) \mid = o_{p}^{\ast }(1),$ in probability.
\end{lemma}

\begin{proof}
The proof follows by arguing as in the proof of Lemma~\ref{lem:hyp:boot} with $\h \bfphi$ replaced by $\h \bfphi^\dag$.
\end{proof}

By Lemmas~\ref{lem:hyp:boot} and~\ref{lem:hyp:boot:2} we obtain that the effect of initialization in the bootstrap data generation is asymptotically negligible.
Hence, in the 
next section we only focus on $\mathcal{U}_n^{*}(\cdot,\h \bfphi^*)$.

\subsection{Main proofs}

This section provides the proofs of the main results stated in the paper.

\begin{proof}[\bf Proof of Lemma~\ref{thm:1}]
First, partition $\cU_n(\cdot, \h \bfphi\,)$ as follows. %
\benr\label{m:bd}
\notag \lefteqn{\cU_n(y,\h \bfphi) - \cU_n(y,\bfphi_0)}\\
\notag &=& n^{-1/2} \s \{Y_i^2/{\it h}_i(\h \bfphi) - Y_i^2/{\it h}_i(\bfphi_0) )\}\mathbb{I}(Y_{i-1} \le y)\\
\notag &=& n^{-1/2} \s \vep_i^2\{{\it h}_i(\bfphi_0)/{\it h}_i(\h \bfphi) - 1\}\mathbb{I}(Y_{i-1} \le y)\\
 &=& n^{-1/2} \s -\vep_i^2\left(\frac{{\it h}_i(\h \bfphi)-{\it h}_i(\bfphi_0)}{{\it h}_i(\bfphi_0)}\right)\mathbb{I}(Y_{i-1} \le y)\\
\notag && + n^{-1/2} \s \vep_i^2\left({\it h}_i(\h \bfphi)-{\it h}_i(\bfphi_0)\right) \left(\frac{1}{{\it h}_i(\bfphi_0)}-\frac{1}{{\it h}_i(\h \bfphi)}\right)\mathbb{I}(Y_{i-1} \le y).
\eenr
Since $\mE(\vep_0^2) =1$ and $n^{1/2}(\h \bfphi - \bfphi_0) = O_p(1)$, by applying Proposition~\ref{prop:A4:A9} and the Ergodic Theorem
to the expansion~\eqref{m:bd}, we obtain that,
uniformly in~$y\in \R$,
\benrr
\cU_n(y,\h \bfphi) &=& \cU_n(y,\bfphi_0) - n^{-1}\s \{ \vep_i^2 n^{1/2}(\h \bfphi - \bfphi_0)^\prime \tau_i(\bfphi_0) \} \mathbb{I}(Y_{i-1} \le y) + o_p(1)\\
&=& \cU_n(y,\bfphi_0) - n^{1/2}(\h \bfphi - \bfphi_0)^\prime \mE[\tau_1(\bfphi_0)\mathbb{I}(Y_{0} \le y)] + o_p(1).
\eenrr
\end{proof}

For the proof of Theorem~\ref{thm:2} we introduce the following additional notation.
For $d \ge 1$, let $\mathbb{C}^d \equiv \cC([-\iny,\iny], \R^d)$ denote the space of continuous functions
from $[-\iny,\iny]$ into $\R^d$.
A sequence of $d$-dimensional stochastic processes (with \textit{cadlag} paths) is said to be $\cC$-tight
if it has associated laws that are tight and whose limit points are concentrated on the
set of continuous paths~$\mathbb{C}^d$.

\begin{proof}[\bf Proof of Theorem~\ref{thm:2}]
From Lemma~\ref{thm:1} and~\eqref{th:expan},
we obtain that, uniformly in $y\in \R$,
\benr\label{hUn:exp}
\notag \lefteqn{\cU_n(y,\h \bfphi) 
 \ = \ \cU_n(y,\bfphi_0)}\\
&& - \, \mE[\tau_1^\prime(\bfphi_0)\mathbb{I}(Y_{0} \le y)]\Si_n^{-1}(\bfphi_0) n^{-1/2}\s (\vep_i^2-1)\tau_i(\bfphi_0)  + o_p(1).
\eenr
By the Ergodic Theorem (e.g., Theorem 2.5.2 of \citealp{Giraitis:Koul:12}) $\Si_n(\bfphi_0) \stackrel{a.s.}{\rightarrow} \Si(\bfphi_0)$, as $n \to \iny$.
Hence, by using the above asymptotic uniform expansion of $\cU_n(\cdot,\h \bfphi)$, we derive that
\benrr
\text{Cov}\{\cU_n(x,\h \bfphi),\cU_n(y,\h \bfphi)\}
&=& K(x,y) + J^\prime(x, \bfphi_0) \mE[M_1(\bfphi_0)M_1^\prime(\bfphi_0)] J^\prime(y, \bfphi_0)\\
&& - J^\prime (x, \bfphi_0)\mE[(\vep_1^2-1)M_1(\bfphi_0)\mathbb{I}(Y_0 \le y)]\\
&& - J^\prime (y, \bfphi_0)\mE[(\vep_1^2-1)M_1(\bfphi_0)\mathbb{I}(Y_0 \le x)] + o(1), \quad x,y \in \R.
\eenrr
Hence, %
$
\text{Cov}\{\cU_n(x,\h \bfphi),\cU_n(y,\h \bfphi)\} = \text{Cov}\{\cU_0(x), \cU_0(y)\} + o(1), \ x,y \in \R, 
$
where $\cU_0$ is the centred Gaussian process in Theorem~\ref{thm:2}.
The convergence of finite dimensional distributions of $\cU_n(\cdot,\h \bfphi)$ can be derived by 
e.g. an application of Theorem~18.3 in \cite{Billingsley:99}.

To show that $\cU_n(y,\h \bfphi)$ is tight, let
$G^{-1}(u):= \inf \{y \in \R : G(y) \ge u\}$ and
\benn
\bar{\cU}_n(u,\bfphi)
:= n^{-1/2} \sum_{i=1}^n \left\{\frac{Y_i^2}{{\it h}_i(\bfphi)} - 1\right\} \mathbb{I}(Y_{i-1} \leq G^{-1}(u)),
\quad u \in [0,1], \ \bfphi \in \Phi.
\eenn
Then, by standard quantile representation, we have that $\cU_n(y,\bfphi) = \bar{\cU}_n(G(y),\bfphi)$ for $y \in \bar \R$.
Let $0\le u_1 \le u \le u_2\le 1$ be fixed but arbitrary. Set
\benrr
&& a_i = \{\vep_i^2-1\}\mathbb{I}(G^{-1}(u_1) < Y_{i-1} \le G^{-1}(u)),\\
&& b_i = \{\vep_i^2-1\}\mathbb{I}(G^{-1}(u) < Y_{i-1} \le G^{-1}(u_2)).
\eenrr
Note that $\mE(a_i)=\mE(b_i)=0$ and $a_ib_i=0$. Further, $\bar{\cU}_n(u,\bfphi_0)-\bar{\cU}_n(u_1,\bfphi_0) =
n^{-1/2} \s a_i$ and $\bar{\cU}_n(u_2,\bfphi_0)-\bar{\cU}_n(u,\bfphi_0) = n^{-1/2} \sum_{j=1}^n b_j$.
Therefore, from Lemma~\ref{lem:billingsley}, we obtain that
\benrr
\lefteqn{\mE[\{\bar{\cU}_n(u,\bfphi_0)-\bar{\cU}_n(u_1,\bfphi_0)\}\{\bar{\cU}_n(u_2,\bfphi_0)-\bar{\cU}_n(u,\bfphi_0)\}]}\\
 &=& n^{-2}\mE\{(\s a_i)^2(\sum_{j=1}^n b_j)^2\}
\ \le \ [n\mE(a_1^2b_1^2)+3n(n-1)\mE(a_1^2)\mE(b_1^2)]/n^2\\
&=& 3n(n-1)n^{-2}[\mE\{\vep_1^2-1\}^2]^2(u-u_1)(u_2-u) \\
&\le& 3(\ka_\vep - 1)^2(u_2-u_1)^2.
\eenrr
Since $u_1$ and $u_2$ are arbitrary, then it follows that $\cU_n(\cdot,\bfphi_0)$ is $\cC$-tight,
e.g. by Theorem~15.7 of \cite{Billingsley:68}.
Since $G$ is continuous, the last term in~\eqref{hUn:exp},
\benrr
&\mE[\tau_1^\prime(\bfphi_0)\mathbb{I}(Y_{0} \le y)]c\Si_n^{-1}(\bfphi_0) n^{-1/2}\s \tau_i(\bfphi_0)\varrho (\vep_i),
\eenrr
 is asymptotically $\cC$-tight,
and hence $\cU_n(\cdot,\h \bfphi_0)$ is also asymptotically $\cC$-tight.
From the latter fact and the convergence of finite dimensional distributions of $\cU_n(\cdot,\h \bfphi)$ to those of $\cU_0(\cdot)$, it follows that
 $\cU_n(\cdot,\h \bfphi) \stackrel{w}{\Rightarrow} \cU_0(\cdot)$ in $\cD(\R)$, where `$\stackrel{w}{\Rightarrow}$' denotes weak convergence of~processes.
\end{proof}

We next state the proof of Theorem~\ref{thm:3}.
In view of Lemmas~\ref{lem:hyp:boot} and~\ref{lem:hyp:boot:2}, and the continuous mapping theorem,
the effect of initialization in the bootstrap data generation is asymptotically negligible.
Therefore, in the next proof, and in the sequel, we do not distinguish between
$\{Y_i^{*(\iny)}, {\it h}_i^{*(\iny)}(\bfphi), \tau_i^{*(\iny)}(\bfphi)\}$ in~\eqref{boot:hyp}, 
and the operational bootstrap process $\{Y_i^*, {\it h}_i^*(\bfphi), \tau_i^*(\bfphi)\}$ in \eqref{boot:op}. 

\begin{proof}[\bf Proof of Theorem~\ref{thm:3}]
By extending the arguments of Lemma~\ref{thm:1} to a  triangular array setup,
we obtain that,
conditional on $\{Y_1, \ldots, Y_n\}$, uniformly over~$y \in \R$,
\benrr
\cU_n^*(y,\h \bfphi^*) &=& \cU_n^*(y,\h \bfphi) - \frac{1}{n}\s (\vep_i^*)^2 n^{1/2}(\h \bfphi^* - \h \bfphi)^\prime \tau_i^*(\h \bfphi) \mathbb{I}(Y_{i-1}^* \le y) + o_{p}^{\ast }(1),\\
&=& \cU_n^*(y,\h \bfphi) - \frac{1}{\sqrt{n}}\s (\h \bfphi^* - \h \bfphi)^\prime \tau_i^*(\h \bfphi) \mathbb{I}(Y_{i-1}^* \le y) + o_{p}^{\ast }(1),
\eenrr
in probability.
Since $\cU_n^*(y,\h \bfphi) = n^{-1/2} \s (\vep_i^* -1)^2 \mathbb{I}(Y_{i-1}^* \le y)$, by using Assumption~\ref{A:1:c:n},
for every $x,y \in \R$, with $x \wedge y := \min (x,y)$, we derive that
\benrr
\cov^*\{\cU_n^*(x,\h \bfphi), \cU_n^*(y,\h \bfphi)\} &=& n^{-1} \s \mE^* (\vep_i^* -1)^2 \mathbb{I}(Y_{i-1}^* \le x \wedge y)\\
&=& (\ka_\vep -1) \mE \mathbb{I}(Y_{i-1} \le x \wedge y) + o_p(1)\\
&=& K(x,y) + o_p(1).
\eenrr
Further, by arguing as for Theorem~\ref{thm:2} in a triangular array, we also obtain that
\benn
\text{Cov}^*\{\cU_n^*(x,\h \bfphi^*),\cU_n^*(y,\h \bfphi^*)\} = K(x,y) + g^*(x,y,\bfphi_0) +  o_p(1), \quad x,y \in \R, %
\eenn
where
\benrr
g^*(x,y,\bfphi_0) &=& J^\prime(x, \bfphi_0) \mE[M_1(\bfphi_0)M_1^\prime(\bfphi_0)] J^\prime(y, \bfphi_0)\\
&& - J^\prime (x, \bfphi_0)\mE[(\vep_1^2-1)M_1(\bfphi_0)\mathbb{I}(Y_0 \le y)]\\
&& - J^\prime (y, \bfphi_0)\mE[(\vep_1^2-1)M_1(\bfphi_0)\mathbb{I}(Y_0 \le x)].
\eenrr
By using the Cramer-Wold device and a CLT for triangular arrays of row-wise independent mean zero
r.v.'s (e.g., Corollary 3.3.1 of \citealp{Hall:Heyde:80})
we obtain that the finite dimensional distributions of $\cU_n^*(\cdot,\h \bfphi^*)$
 converge to those of $\cU_0$, in probability,
where $\cU_0$ is the centred Gaussian process in Theorem~\ref{thm:2}.
Further, by extending the arguments of Theorem~\ref{thm:2} to a triangular array we also obtain that $\cU_n^*(\cdot,\h \bfphi^*)$ is asymptotically $\cC$-tight.
Hence the part~1 follows.
The part~2 follows from an application of the continuous mapping theorem.
Since $\cU_n^*(\cdot,\h \bfphi^*)$ converges weakly, and $n^{1/2}(\h \bfphi^* - \h \bfphi) = O_{p}^{\ast }(1),$ in probability,
part~3 also holds for the KS and CvM functional~forms.
\end{proof}

\begin{proof}[\bf Proof of Theorem~\ref{thm:power}]
If some components of $\bfphi_0^*$ lie on the boundary of the parameter space; i.e.,
$\phi_{0i}^* = 0$ for some $i = 2, \ldots, p_1+p_2+1$,  then the proof follows from arguing as in the proof of Theorem~\ref{thm:power:2}.
Hence, here we only consider the case $\bfphi_0^*$ is in the interior of~$\Phi$.
Since 
Assumptions~\ref{A:1}--\ref{A:3:c} hold, and $(\h \bfphi, \check F_n) \overset{p}{\rightarrow } (\bfphi_0^*, F_0^*)$,
except that $(\bfphi_0^*, F_0^*)$ is the pseudo-true value under $\mathsf{H}_1$,
by arguing as in the proof of Theorem~\ref{thm:3},
conditional on $\{Y_1, \ldots, Y_n\}$, the process $\cU_n^*(\cdot,\h \bfphi^*)$
converges weakly to the centred Gaussian process $\cU_0^\dag(\cdot)$ specified by the covariance~kernel
\benrr
\text{\em Cov} \{\cU_0^\dag(x), \cU_0^\dag(y)\} &=& \mE (\{{F_0^*}^{-1}(U_i)\}^2-1)^2 \mathbb{I}(Y_{i-1} \le x\wedge y)\\
&& + J^\prime(x, \bfphi_0^*) \mE[V_i(\bfphi_0)V_i^\prime(\bfphi_0^*)] J^\prime(y, \bfphi_0^*)\\
&& - J^\prime (x, \bfphi_0^*)\mE[(\{{F_0^*}^{-1}(U_i)\}^2-1)V_i(\bfphi_0^*)\mathbb{I}(Y_{i-1} \le y)]\\
&& - J^\prime (y, \bfphi_0^*)\mE[(\{{F_0^*}^{-1}(U_i)\}^2-1)V_i(\bfphi_0^*)\mathbb{I}(Y_{i-1} \le x)],
\eenrr
in probability, where $\bfU = \{U_i, i \in \mathbb{Z}\}$ are i.i.d. uniform(0,1) random variables,
\benn
V_i(\bfphi) := - \Si^{-1}(\bfphi) [1-\{{F_0^*}^{-1}(U_i)\}^2]\tau_i(\bfphi), \quad \bfphi \in \Phi.
\eenn
Therefore, it suffices to show that $n^{-1/2}|\cU_n(y,\h \bfphi)|=O_p(1)$, for some $y\in \R$ satisfying Assumption~\ref{A:power}. Fix such a $y$.
Under $\mathsf{H}_1$, $\vep_i=Y_i/\sqrt{{\it h}_i}$, where ${\it h}_i = \mE[Y_i^2 \mid \cH_{i-1}]$, $i \in \Z$. Therefore, we have that
\benr\label{uny0}
n^{-1/2}|\cU_n(y,\h \bfphi)| &\le& \big | n^{-1} \s \vep_i^2 {\it h}_i\big[{\it h}_i^{-1}(\h \bfphi) - {\it h}_i^{-1}(\bfphi_0^*)\big]
\mathbb{I}(Y_{i-1} \le y) \big | \nn\\
&&+ \big | n^{-1} \s \big \{\vep_i^2 \big({\it h}_i/{\it h}_i(\bfphi_0^*)\big) - 1 \big \} \mathbb{I}(Y_{i-1} \le y) \big |.
\eenr
Lemma~\ref{lem:A4:A9} holds under Assumptions 
\ref{A:1}--\ref{A:3:c},
with $\bfphi_0^*$ denoting the pseudo-true value under $\mathsf{H}_1$.
Therefore, on the set $n^{1/2}\|\h \bfphi - \bfphi_0^*\|\le K$,
the first term on the right hand side
of (\r{uny0}) is
bounded from the above by
$
\om_L^{-1} n^{-1} \s Y_i^2 \big |(\h \bfphi - \bfphi_0^*)^\prime \tau_i(\bfphi_0^*)\big | + o_p(n^{-1/2}) =o_p(1),
$
by the Ergodic Theorem,  and because  $\h \bfphi \overset{p}{\rightarrow } \bfphi_0^*$.
Since $n^{1/2}(\h \bfphi - \bfphi_0^*) = O_p(1)$, then by using an extended Glivenko-Cantelli type
argument, we obtain that
\benrr\label{uny}
n^{-1/2}|\cU_n(y,\h \bfphi)| \ = \ |\mE\big([{\it h}_1/{\it h}_1(y,\bfphi_0^*) - 1]\mathbb{I}(Y_0\le y)\big)| +  o_p(1),
\ \mb{under $\mathsf{H}_1$}. \
\eenrr
Hence, the proof follows from Assumption~\ref{A:power}.
\end{proof}

\begin{proof}[\bf Proof of Theorem~\ref{thm:6}]
Irrespective of whether $\bfphi_0$ is in the interior or on the boundary of the parameter space,
by arguing as in the proof of Lemma~\ref{thm:1}, we obtain~that
\ben\label{check}
\sup_{y\in \R} \mid \cU_n(y,\h \bfphi) - \cU_n(y,\bfphi_0)  + n^{1/2}(\h \bfphi - \bfphi_0)^\prime J(y,\bfphi_0) \mid = o_p(1),
\een
with $\cU_n(\cdot,\bfphi_0)$ converging weakly to a centred Gaussian process with covariance kernel
$K(x,y) = (\ka_\vep-1) G(x\wedge y), \  x, y \in \R$.

To establish the validity of the bootstrap tests we first consider the case $\bfphi_0$ is in the interior of the parameter space.
Since $\h \bfphi$ is the QMLE in~\eqref{lth}, 
\ben\label{check2}
n^{1/2}(\h \bfphi-\bfphi_0) = Z_n + o_p(1),
\een
where
\benn
Z_n := -\Si_n^{-1}(\bfphi_0)n^{-1/2}\s (1-\vep_i^2)\tau_i(\bfphi_0), \quad \Si_n(\bfphi) := n^{-1}\s \tau_i(\bfphi) \tau_i(\bfphi)^\prime,
\eenn
and hence, with the asymptotic uniform expansion of $\cU_n(\cdot, \h \bfphi)$ in~\eqref{check}, it follows as in the proof of Theorem~\ref{thm:2} that
$\cU_n(\cdot,\h \bfphi)$ converges weakly to $\cU_0(\cdot)$
in $\cD(\R)$,  where $\cU_0$ is the centred Gaussian process given in Theorem~\ref{thm:2}.

In the method of bootstrap data generation outlined in Section~\ref{sec:boot:new}
the transformed estimator $\h \bfphi^\dag$
plays the role of the true parameter $\bfphi_0$;
recall that
$\h \bfphi^\dag = \h \bfphi_n^\dag = (\h \bfphi_{n1}^\dag, \ldots, \h \bfphi_{n(1+p_1+p_2)}^\dag)^\prime$ where
$
\h \bfphi^\dag_{ni} := \h \bfphi_{ni} \mathbb{I}(\h \bfphi_{ni} > c_n),$ $i= 1,2, \ldots, 1+p_1+p_2,
$
and ($c_n$) is a non-random sequence with $c_n \to 0$ and $n^{1/2}c_n \to \iny$ as $n \to \iny$.
Let $A_{ni} = \{\h \bfphi_{ni} > c_n\}, \ i= 1,2, \ldots, 1+p_1+p_2$.
Since $\bfphi_0$ is an interior point, we have $\bfphi_{0j} > 0$, $j= 1, \ldots, 1+p_1+p_2$.
Further, as $c_n$ converges to $0$ at a rate slower than $n^{-1/2}$ and $n^{1/2}(\h \bfphi - \bfphi_0)=O_p(1)$, we obtain that
$P(\cap_{i=1}^{1+p_1+p_2} A_{ni}) \ra 1$ as $n \to \iny$.
Since $\h \bfphi^\dag = \h \bfphi$ on the set $\cap_{i=1}^{1+p_1+p_2} A_{ni}$, then
the asymptotic validity of the bootstrap tests follows from the same arguments used in the proof of Theorem~\ref{thm:3}.

Next, we consider the validity of the bootstrap tests for the case some components of $\bfphi_0$ lie on the boundary of the parameter space; i.e.,
$\phi_{0i} = 0$ for some $i = 2, \ldots, p_1+p_2+1$.
Since $\bfphi_0$ is not an interior point, in this case, the limiting behaviour of $n^{1/2}(\h \bfphi-\bfphi_0)$ is not
linear as in~\eqref{check2}, and as in the proof of Theorem 2 of \cite{Francq:Zakoian:07}, we obtain
 \ben\label{lan}
 n^{1/2}(\h \bfphi-\bfphi_0) = \la_n^\Lambda + o_p(1), \quad \la_n^\Lambda := \arg\inf_{\la \in \Lambda} (Z_n - \la)^\prime \Si_n(\bfphi_0) (Z_n - \la).
 \een
The vector $\la_n^\Lambda$ is the orthogonal projection of $Z_n$ on the convex set $\Lambda$ for the inner product $\langle x, y \rangle := x^\prime \Si_n(\bfphi_0) y$,
and it is characterized by $\la_n^\Lambda \in \Lambda$, $\langle Z_n - \la_n^\Lambda, \la_n^\Lambda - \la \rangle \ge 0$, $\forall \la \in \Lambda$; see e.g. Lemma 1.1 in \cite{Zarantonello:71}.
Thus, by arguing as in the proof of Theorem 2 in \cite{Francq:Zakoian:07} we obtain that
$n^{1/2}(\h \bfphi - \bfphi_0)  \stackrel{d}{\rightarrow} \la^\Lambda := \arg\inf_{\la \in \Lambda} (\la - Z)^\prime \Si(\bfphi_0) (\la - Z).$
Therefore, for a continuous $G$, the term $n^{1/2}(\h \bfphi - \bfphi_0)^\prime J(y,\bfphi_0)$ in~\eqref{check} is asymptotically $\cC$-tight,
as is $\cU_n(\cdot,\bfphi_0)$ by Lemma~\ref{thm:1} and Theorem~\ref{thm:2}. %

Next, consider the bootstrap data generation as outlined in Section~\ref{sec:boot:new}.
Since $\h \bfphi^\dag$ converges to $\bfphi_0$, a.s., by a triangular array version of the proof of Lemma~\ref{thm:1}
replacing $\h \bfphi$ and $\bfphi_0$ by $\h \bfphi^*$ and $\h \bfphi^\dag$, respectively, we obtain that conditional on $\{Y_1, \ldots, Y_n\}$, uniformly in~$y \in \R$,
\ben\label{hU-U:boot}
\cU_n^*(y,\h \bfphi^*) =\cU_n^*(y,\h \bfphi^\dag)  - n^{1/2}(\h \bfphi^* - \h \bfphi^\dag)^\prime J^*(y,\h \bfphi^\dag) + o_{p}^{\ast }(1),
\een
in probability, where
\benn
J^*(y,\bfphi) = \mE^*[\tau_1^*(\bfphi)\mathbb{I}(Y_{0}^* \le y)], \quad \tau_i^*(\bfphi)
:= \frac{(\partial/\partial \bfphi) {\it h}_i^*(\bfphi)}{{\it h}_i^*(\bfphi)}, \quad \bfphi \in \Phi.
\eenn
Since $\cU_n^*(y,\h \bfphi^\dag) = n^{-1/2} \s (\vep_i^* -1)^2 \mathbb{I}(Y_{i-1}^* \le y)$, by using Assumption~\ref{A:1:c:n},
for every $x,y \in \R$, with $x \wedge y := \min (x,y)$, we obtain that
\benr\label{hU-U:cov}
\notag \cov^*\{\cU_n^*(x,\h \bfphi^\dag), \cU_n^*(y,\h \bfphi^\dag)\} &=& n^{-1} \s \mE^* (\vep_i^* -1)^2 \mathbb{I}(Y_{i-1}^* \le x \wedge y)\\
\notag &=& (\ka_\vep -1) \mE \mathbb{I}(Y_{i-1} \le x \wedge y) + o_p(1)\\
&=& \cov\{\cU_n(x,\bfphi_0), \cU_n(y,\bfphi_0)\} + o_p(1). \quad
\eenr
Further, it follows from Assumption~\ref{A:1:c:n} that $J^*(y,\bfphi) = \mE[\tau_1(\bfphi)\mathbb{I}(Y_{0} \le y)] + o_p(1)$.

Therefore, in order to establish that the conditional weak limit of $\cU_n^*(\cdot,\h \bfphi^*)$ is the same as that of $\cU_n(\cdot,\h \bfphi)$, in probability,
we need to first show that the conditional limiting distribution of $n^{1/2}(\h \bfphi^* - \h \bfphi^\dag)$
is the same as that of $n^{1/2}(\h \bfphi - \bfphi_0)$, in probability.
To this end, it suffices to show that, conditional on $\{Y_1, \ldots, Y_n\}$,
\ben\label{boot:con:phi}
n^{1/2}(\h \bfphi^* - \h \bfphi^\dag) = \la_n^\Lambda + o_{p}^{\ast }(1), 
\text{ in probability.}
\een
In order to obtain \eqref{boot:con:phi} we consider a triangular array version of the proof of~\eqref{lan}.

In the proof of~\eqref{lan} in \cite{Francq:Zakoian:07}, first
$\la_n^\Lambda$ is represented as the orthogonal projection of $Z_n$ on the convex set $\Lambda$, for the inner product $\langle x, y \rangle := x^\prime \Si_n(\bfphi_0) y$,
and then this projection is approximated by that of $Z_n$ on the set $n^{1/2}(\Phi-\bfphi_0)$.
Since $\Phi$ contains a hypercube which includes $\bfphi_0$, see~\eqref{hcube},  the set $n^{1/2}(\Phi-\bfphi_0)$ increases to $\Lambda$ as $n \to \iny$.
This plays a key role in the proof of~\eqref{lan}.
Recall that,
\benn
\Lambda = \Lambda_1 \times \Lambda_2 \times \cdots \times \Lambda_{p_1+p_2+1},
\eenn
where $\Lambda_1 = \R$, and for each $i = 2, \ldots, p_1+p_2+1$,
denoting $\bfphi_0 = (\bfphi_{01}, \ldots, \bfphi_{0(1+p_1+p_2)})^\prime$,
$\Lambda_i = \R$ if $\bfphi_{0i} \neq 0$ and $\Lambda_i = [0,\iny)$ if $\bfphi_{0i} = 0$.
In order to extend the proof of~\eqref{lan} to the triangular array setup of the bootstrap data generation,
we need to replace $\bfphi_0$ by the bootstrap true parameter $\h \bfphi^\dag$.
However, to ensure that $n^{1/2}(\Phi-\h \bfphi^\dag)$ increases to $\Lambda$, 
we need to show that $\h \bfphi^\dag$ satisfies two important conditions.
First, 
to allow the hypercube in~\eqref{hcube} to contain $\h \bfphi^\dag$ with probability converging to one, we need to have
\ben\label{pdag:con}
\h \bfphi^\dag \ra \bfphi_0 \ \text{in probability as $n \to \iny$}.
\een
Further, $\h \bfphi^\dag$ should satisfy the following rate of consistency property.
\ben\label{pdag:con:2}
n^{1/2}(\h \bfphi_{ni}^\dag - \bfphi_{0i}) =
\left\{
  \begin{array}{ll}
    o_p(1), & \hbox{if $\bfphi_{0i}=0$} \\
    O_p(1), & \hbox{if $\bfphi_{0i}>0$}
  \end{array}
\right., \quad
i= 1,2, \ldots, 1+p_1+p_2.
\een
Since at least one component of $\bfphi_0$ is zero, the rate of consistency \eqref{pdag:con:2} ensures that 
$
n^{1/2}(\Phi-\h \bfphi^\dag) = n^{1/2}(\Phi - \bfphi_0) - n^{1/2}(\bfphi^\dag - \bfphi_0)
$
converges to $\Lambda$ in probability.

The consistency of $\h \bfphi^\dag$ follows from that of $\h \bfphi$, and hence~\eqref{pdag:con} holds.
Since $c_n$ converges to $0$ at a rate slower than $n^{-1/2}$, \eqref{pdag:con:2} follows by arguing as in the proof of Lemma~1 in \cite{Cavaliere:19}.
Hence, by a triangular array extension of the proof~of Lemma~\ref{thm:1g}, under Assumption~\ref{qml:7}, %
we obtain that \eqref{boot:con:phi} holds,
for example by arguing as in the proof of Proposition 3.2 in \cite{Hidalgo:Zaffaroni:07}; 
see also the discussion under Assumption~E2 in \cite{Andrews:97}.
Therefore, from \eqref{check}--\eqref{hU-U:boot}, and the asymptotic tightness of $\cU_n^*(\cdot,\h \bfphi^\dag)$,
we obtain that the conditional weak limit of
$\cU_n^*(\cdot,\h \bfphi^*)$ is the same as that~of $\cU_n(\cdot,\h \bfphi)$, in probability.
Hence, the continuous mapping theorem yields that the bootstrap test~\eqref{eq:boot:tst:2} based on $T_j$ is asymptotically valid ($j = 1, 2$).
\end{proof}

\begin{proof}[\bf Proof of Theorem~\ref{thm:power:2}]
If $\bfphi_0^*$ is an interior point, then the proof follows from Theorem~\ref{thm:power}.
Therefore, we only consider the case some components of $\bfphi_0^*$ lie on the boundary of the parameter space; i.e.,
$\phi_{0i}^* = 0$ for some $i = 2, \ldots, p_1+p_2+1$.
Since Assumptions~\ref{A:1}, \ref{A:1:c}, \ref{A:1:c:n}, \ref{A:3:c} and~\ref{qml:7} continue to hold,
although $(\bfphi_0^*,F_0^*)$ is the pseudo-true value under $\mathsf{H}_1$, by arguing as in the proof of Theorem~\ref{thm:6}, for every $y \in \R$,
we have that $\cU_n^*(y,\h \bfphi^\dag) = O_{p}^{\ast }(1)$, in probability. %
Therefore, it suffices to show that $n^{-1/2}|\cU_n(y,\h \bfphi)|=O_p(1)$,
for some $y\in \R$ satisfying $\mE[\{{\it h}_1/{\it h}_1(\bfphi_0^*) - 1\}\mathbb{I}(Y_0 \le y)] \neq 0$. Fix such a~$y$.
Under $\mathsf{H}_1$, $\vep_i=Y_i/\sqrt{{\it h}_i}$, where ${\it h}_i = \mE[Y_i^2 \mid \cH_{i-1}]$. %
Hence, by using Assumptions~\ref{A:1}, \ref{A:1:c}, \ref{A:1:c:n}, and~\ref{A:3:c}, 
and arguing as in the proof of Lemma~\ref{lem:A4:A9},
on the set $n^{1/2}\|\h \bfphi - \bfphi_0^*\|\le C$, where $0<C<\iny$, we obtain that
\benrr
\lefteqn{\big | n^{-1} \s Y_i^2\big[{\it h}_i^{-1}(\h \bfphi) - {\it h}_i^{-1}(\bfphi_0^*)\big]\mathbb{I}(Y_{i-1} \le y) \big |}\\
 &\le& \om_L^{-1} n^{-1} \s Y_i^2 \big |(\h \bfphi - \bfphi_0^*)^\prime \tau_i(\bfphi_0^*)\big | + o_p(n^{-1/2}).
\eenrr
By the Ergodic Theorem,  and because  $\h \bfphi \overset{p}{\rightarrow } \bfphi_0^*$, the sum in the above upper bound is $o_p(1)$,
and hence
$n^{-1/2}|\cU_n(y,\h \bfphi)|$ is bounded from the above by
$| n^{-1} \s \big \{\vep_i^2 \big({\it h}_i/{\it h}_i(\bfphi_0^*)\big) - 1 \big \} \mathbb{I}(Y_{i-1} \le y) |$
up to a term of order $o_p(1)$.
Since $n^{1/2}(\h \bfphi - \bfphi_0^*) = O_p(1)$, then the proof follows by an extended Glivenko-Cantelli type
argument as in the proof of Theorem~\ref{thm:power}.
\end{proof}

\end{appendix}

\bibliographystyle{natbib}

\begin{thebibliography}{}

\bibitem[Agosto {\em et~al.}(2016)]{AGOSTO:16}
Agosto, A., Cavaliere, G., Kristensen, D., and Rahbek, A. (2016).
\newblock Modeling corporate defaults: Poisson autoregressions with exogenous
  covariates (parx).
\newblock {\em Journal of Empirical Finance}, {\bf 38}, 640--663.

\bibitem[Andrews(1997)]{Andrews:97}
Andrews, D. W.~K. (1997).
\newblock A conditional {K}olmogorov test.
\newblock {\em Econometrica}, {\bf 65}(5), 1097--1128.

\bibitem[Andrews(2001)]{Andrews:01}
Andrews, D. W.~K. (2001).
\newblock Testing when a parameter is on the boundary of the maintained
  hypothesis.
\newblock {\em Econometrica}, {\bf 69}(3), 683--734.

\bibitem[Bai(2003)]{Bai:03}
Bai, J. (2003).
\newblock Testing parametric conditional distributions of dynamic models.
\newblock {\em The Review of Economics and Statistics}, {\bf 85}(3), 531--549.

\bibitem[Balakrishna {\em et~al.}(2019)]{Bala:Koul:19}
Balakrishna, N., Koul, H.~L., Sakhanenko, L., and Ossiander, M. (2019).
\newblock Fitting a {$p$}th order parametric generalized linear autoregressive
  multiplicative error model.
\newblock {\em Sankhya B}, {\bf 81}(1, suppl.), S103--S122.

\bibitem[Berkes and Horv{\'a}th(2004)]{Berkes:Horvath:2004}
Berkes, I. and Horv{\'a}th, L. (2004).
\newblock The efficiency of the estimators of the parameters in {GARCH}
  processes.
\newblock {\em The Annals of Statistics}, {\bf 32}(2), 633--655.

\bibitem[Berkes {\em et~al.}(2003)]{Berkes:etal:03}
Berkes, I., Horv{\'a}th, L., and Kokoszka, P. (2003).
\newblock G{ARCH} processes: structure and estimation.
\newblock {\em Bernoulli}, {\bf 9}(2), 201--227.

\bibitem[Bickel and Freedman(1981)]{Bickel:Freedman:81}
Bickel, P.~J. and Freedman, D.~A. (1981).
\newblock Some asymptotic theory for the bootstrap.
\newblock {\em The Annals of Statistics}, {\bf 9}(6), 1196--1217.

\bibitem[Bickel {\em et~al.}(1998)]{Bickel:etal:98}
Bickel, P.~J., Klaassen, C. A.~J., Ritov, Y., and Wellner, J.~A. (1998).
\newblock {\em Efficient and adaptive estimation for semiparametric models}.
\newblock Springer-Verlag, New York.
\newblock Reprint of the 1993 original.

\bibitem[Billingsley(1968)]{Billingsley:68}
Billingsley, P. (1968).
\newblock {\em Convergence of probability measures}.
\newblock John Wiley \& Sons Inc., New York.

\bibitem[Billingsley(1999)]{Billingsley:99}
Billingsley, P. (1999).
\newblock {\em Convergence of probability measures}.
\newblock Wiley Series in Probability and Statistics: Probability and
  Statistics. John Wiley \& Sons Inc., New York, second edition.
\newblock A Wiley-Interscience Publication.

\bibitem[Bollerslev(1986)]{Bollerslev:86}
Bollerslev, T. (1986).
\newblock Generalized autoregressive conditional heteroskedasticity.
\newblock {\em Journal of Econometrics}, {\bf 31}(3), 307--327.

\bibitem[Bougerol and Picard(1992a)]{Bougerol:Picard:92}
Bougerol, P. and Picard, N. (1992a).
\newblock Stationarity of {GARCH} processes and of some nonnegative time
  series.
\newblock {\em Journal of Econometrics}, {\bf 52}(1-2), 115--127.

\bibitem[Bougerol and Picard(1992b)]{bougerol1992}
Bougerol, P. and Picard, N. (1992b).
\newblock Strict stationarity of generalized autoregressive processes.
\newblock {\em The Annals of Probability}, {\bf 20}(4), 1714--1730.

\bibitem[Cavaliere {\em et~al.}(2021)]{Cavaliere:19}
Cavaliere, G., Nielsen, H.~B., Pedersen, R.~S., and Rahbek, A. (2021).
\newblock {Bootstrap Inference On The Boundary Of The Parameter Space With
  Application To Conditional Volatility Models}.
\newblock {\em Journal of Econometrics}, {\bf in press}.

\bibitem[Chan {\em et~al.}(2014)]{Chan:Tong:14}
Chan, K.-S., Li, D., Ling, S., and Tong, H. (2014).
\newblock On conditionally heteroscedastic {AR} models with thresholds.
\newblock {\em Statistica Sinica}, {\bf 24}(2), 625--652.

\bibitem[Chatterjee and Lahiri(2011)]{Chatterjee:Lahiri:11}
Chatterjee, A. and Lahiri, S.~N. (2011).
\newblock Bootstrapping lasso estimators.
\newblock {\em Journal of the American Statistical Association}, {\bf
  106}(494), 608--625.

\bibitem[Chen and An(1998)]{Chen:Hong:98}
Chen, M. and An, H.~Z. (1998).
\newblock A note on the stationarity and the existence of moments of the
  {GARCH} model.
\newblock {\em Statist. Sinica}, {\bf 8}(2), 505--510.

\bibitem[D'Agostino and Stephens(1986)]{Stephens:86}
D'Agostino, R.~B. and Stephens, M.~A., editors (1986).
\newblock {\em Goodness-of-fit techniques}, volume~68 of {\em Statistics:
  Textbooks and Monographs}.
\newblock Marcel Dekker Inc., New York.

\bibitem[Escanciano(2008)]{Escanciano:2008}
Escanciano, J.~C. (2008).
\newblock Joint and marginal specification tests for conditional mean and
  variance models.
\newblock {\em Journal of Econometrics}, {\bf 143}(1), 74--87.

\bibitem[Escanciano(2010)]{Escanciano:2010}
Escanciano, J.~C. (2010).
\newblock Asymptotic distribution-free diagnostic tests for heteroskedastic
  time series models.
\newblock {\em Econometric Theory}, {\bf 26}(3), 744--773.

\bibitem[Escanciano {\em et~al.}(2018)]{Escanciano:etal:18}
Escanciano, J.~C., Pardo-Fern\'andez, J.~C., and Keilegom, I.~V. (2018).
\newblock Asymptotic distribution-free tests for semiparametric regressions
  with dependent data.
\newblock {\em The Annals of Statistics}, (in press).

\bibitem[Fernandes and Grammig(2005)]{Fernandes:Gramg:05}
Fernandes, M. and Grammig, J. (2005).
\newblock Nonparametric specification tests for conditional duration models.
\newblock {\em Journal of Econometrics}, {\bf 127}(1), 35--68.

\bibitem[Francq and Zakoian(2007)]{Francq:Zakoian:07}
Francq, C. and Zakoian, J.-M. (2007).
\newblock Quasi-maximum likelihood estimation in {GARCH} processes when some
  coefficients are equal to zero.
\newblock {\em Stochastic Processes and their Applications}, {\bf 117}(9),
  1265--1284.

\bibitem[Francq and Zako{\"{\i}}an(2010)]{Francq:2010}
Francq, C. and Zako{\"{\i}}an, J.-M. (2010).
\newblock {\em {GARCH} models: structure, statistical inference and financial
  applications}.
\newblock Wiley. John Wiley \& Sons Ltd.

\bibitem[Giacomini {\em et~al.}(2013)]{Giacomini:etal:07}
Giacomini, R., Politis, D.~N., and White, H. (2013).
\newblock A warp-speed method for conducting {M}onte {C}arlo experiments
  involving bootstrap estimators.
\newblock {\em Econometric Theory}, {\bf 29}(3), 567--589.

\bibitem[Giraitis {\em et~al.}(2012)]{Giraitis:Koul:12}
Giraitis, L., Koul, H.~L., and Surgailis, D. (2012).
\newblock {\em Large sample inference for long memory processes}.
\newblock Imperial College Press, London.

\bibitem[Glosten {\em et~al.}(1993)]{Glosten:GJR:93}
Glosten, L.~R., Jagannathan, R., and Runkle, D.~E. (1993).
\newblock On the relation between the expected value and the volatility of the
  nominal excess return on stocks.
\newblock {\em Journal of Finance}, {\bf 48}(5), 1779--1801.

\bibitem[Grenander and Rosenblatt(1957)]{Grenander:Rosenblatt:57}
Grenander, U. and Rosenblatt, M. (1957).
\newblock {\em Statistical analysis of stationary time series}.
\newblock John Wiley \& Sons, New York; Almqvist \& Wiksell, Stockholm.

\bibitem[Hall and Heyde(1980)]{Hall:Heyde:80}
Hall, P. and Heyde, C.~C. (1980).
\newblock {\em Martingale limit theory and its application}.
\newblock Academic Press Inc. [Harcourt Brace Jovanovich Publishers], New York.
\newblock Probability and Mathematical Statistics.

\bibitem[Hidalgo and Zaffaroni(2007)]{Hidalgo:Zaffaroni:07}
Hidalgo, J. and Zaffaroni, P. (2007).
\newblock A goodness-of-fit test for {ARCH($\infty$)} models.
\newblock {\em Journal of Econometrics}, {\bf 141}(2), 835--875.

\bibitem[Jensen and Rahbek(2004)]{Jensen:Rahbek:04}
Jensen, S. r.~T. and Rahbek, A. (2004).
\newblock Asymptotic inference for nonstationary {GARCH}.
\newblock {\em Econometric Theory}, {\bf 20}(6), 1203--1226.

\bibitem[Khmaladze(1981)]{khm:81}
Khmaladze, {\`E}.~V. (1981).
\newblock A martingale approach in the theory of goodness-of-fit tests.
\newblock {\em Theory of Probability and Its Applications}, {\bf 26}(2),
  240--257.

\bibitem[Kolmogorov(1933)]{Kolmogorov:33}
Kolmogorov, A. (1933).
\newblock Sulla determinazione empirica di una lgge di distribuzione.
\newblock {\em Inst. Ital. Attuari, Giorn.}, {\bf 4}, 83--91.

\bibitem[Koul and Ling(2006)]{Koul:06}
Koul, H.~L. and Ling, S. (2006).
\newblock Fitting an error distribution in some heteroscedastic time series
  models.
\newblock {\em The Annals of Statistics}, {\bf 34}(2), 994--1012.

\bibitem[Koul and Stute(1999)]{Koul:Stute:99}
Koul, H.~L. and Stute, W. (1999).
\newblock Nonparametric model checks for time series.
\newblock {\em The Annals of Statistics}, {\bf 27}(1), 204--236.

\bibitem[Koul {\em et~al.}(2012)]{Koul:Indee:12}
Koul, H.~L., Perera, I., and Silvapulle, M.~J. (2012).
\newblock Lack-of-fit testing of the conditional mean function in a class of
  {Markov} multiplicative error models.
\newblock {\em Econometric Theory}, {\bf 28}(6), 1283--1312.

\bibitem[Nelson(1990)]{Nelson:90}
Nelson, D.~B. (1990).
\newblock Stationarity and persistence in the {GARCH{$(1,1)$}} model.
\newblock {\em Econometric Theory}, {\bf 6}(3), 318--334.

\bibitem[Perera and Koul(2017)]{Indee:Hira:15}
Perera, I. and Koul, H.~L. (2017).
\newblock Fitting a two phase threshold multiplicative error model.
\newblock {\em Journal of Econometrics}, {\bf 197}(2), 348--367.

\bibitem[Perera and Silvapulle(2017)]{Perera:Silvapulle:15a}
Perera, I. and Silvapulle, M.~J. (2017).
\newblock Specification tests for multiplicative error models.
\newblock {\em Econometric Theory}, {\bf 33}(2), 413--438.

\bibitem[Perera and Silvapulle(2020)]{Perera:Silvapulle:21}
Perera, I. and Silvapulle, M.~J. (2020).
\newblock Specification tests for dynamic conditional distribution models.
\newblock {\em Discussion paper}.

\bibitem[Perera and Silvapulle(2021)]{Perera:Silvapulle:20}
Perera, I. and Silvapulle, M.~J. (2021).
\newblock Bootstrap based probability forecasting in multiplicative error
  models.
\newblock {\em Journal of Econometrics}, {\bf 221}(1), 1--24.

\bibitem[Perera {\em et~al.}(2016)]{Perera:Silvapulle:14}
Perera, I., Hidalgo, J., and Silvapulle, M.~J. (2016).
\newblock A goodness-of-fit test for a class of autoregressive conditional
  duration models.
\newblock {\em Econometric Reviews}, {\bf 35}(6), 1111--1141.

\bibitem[Straumann and Mikosch(2006)]{Straumann:Mikosch:06}
Straumann, D. and Mikosch, T. (2006).
\newblock Quasi-maximum-likelihood estimation in conditionally heteroscedastic
  time series: a stochastic recurrence equations approach.
\newblock {\em The Annals of Statistics}, {\bf 34}(5), 2449--2495.

\bibitem[Stute(1997)]{Stute:97}
Stute, W. (1997).
\newblock Nonparametric model checks for regression.
\newblock {\em The Annals of Statistics}, {\bf 25}(2), 613--641.

\bibitem[Tsay and Chen(2018)]{tsay2018}
Tsay, R. and Chen, R. (2018).
\newblock {\em Nonlinear Time Series Analysis}.
\newblock Wiley Series in Probability and Statistics. Wiley.

\bibitem[Tsay(2013)]{Tsay:12}
Tsay, R.~S. (2013).
\newblock {\em An introduction to analysis of financial data with {R}}.
\newblock Wiley Series in Probability and Statistics. John Wiley \& Sons, Inc.,
  Hoboken, NJ.

\bibitem[von Neumann(1941)]{Neumann:41}
von Neumann, J. (1941).
\newblock Distribution of the ratio of the mean square successive difference to
  the variance.
\newblock {\em Annals of Mathematical Statistics}, {\bf 12}, 367--395.

\bibitem[Zarantonello(1971)]{Zarantonello:71}
Zarantonello, E.~H. (1971).
\newblock Projections on convex sets in {H}ilbert space and spectral theory.
  {I}. {P}rojections on convex sets.
\newblock In {\em Contributions to nonlinear functional analysis ({P}roc.
  {S}ympos., {M}ath. {R}es. {C}enter, {U}niv. {W}isconsin, {M}adison, {W}is.,
  1971)}, pages 237--341.

\end{thebibliography}

\end{document}